\documentclass[ALICE,manyauthors]{cernphprep}
\usepackage[comma,square,numbers,sort&compress]{natbib}

\usepackage{lineno}
\usepackage{xspace}
\usepackage{hyperref}
\usepackage[T1]{fontenc}
\usepackage{orcidlink}
\usepackage{changepage}
\usepackage{multirow}

\begin{document}
\newcommand\TBD[1]{\textbf{\color{red}TBD: {#1}}}
\newcommand{\pel}{\ensuremath{\rm e}\xspace}
\newcommand{\pmu}{\ensuremath{\mu}\xspace}
\newcommand{\pde}{\ensuremath{\rm d}\xspace}
\newcommand{\dbar}{\ensuremath{\rm \bar{d}}\xspace}
\newcommand{\ppip}{\ensuremath{\pi^{+}}\xspace}
\newcommand{\ppim}{\ensuremath{\pi^{-}}\xspace}
\newcommand{\ppi}{\ensuremath{\pi}\xspace}
\newcommand{\ppipm}{\ensuremath{\pi^{\pm}}\xspace}
\newcommand{\ppis}{\ensuremath{\ppip + \ppim}\xspace}
\newcommand{\pkap}{\ensuremath{{\rm K}^{+}}\xspace}
\newcommand{\pkam}{\ensuremath{{\rm K}^{-}}\xspace}
\newcommand{\pka}{\ensuremath{{\rm K}}\xspace}
\newcommand{\pkapm}{\ensuremath{{\rm K}^{\pm}}\xspace}

\newcommand{\Xis}{\ensuremath{\rm \Xi}\xspace}
\newcommand{\xim}{\ensuremath{{\rm \Xi}^{-}}\xspace}
\newcommand{\xip}{\ensuremath{{\rm \overline{\Xi}}^{+}}\xspace}


\newcommand{\Oms}{\ensuremath{\rm \Omega}\xspace}
\newcommand{\omm}{\ensuremath{{\rm \Omega}^{-}}\xspace}
\newcommand{\omp}{\ensuremath{{\rm \overline{\Omega}}^{+}}\xspace}

\newcommand{\pkastar}{\ensuremath{{\rm K}^{*}}\xspace}
\newcommand{\pkastarm}{\ensuremath{\overline{\pkastar}}\xspace}
\newcommand{\pkazero}{\ensuremath{{\rm K}^{0}_{\rm S}}\xspace}
\newcommand{\pprp}{\ensuremath{{\rm p}}\xspace}
\newcommand{\pprm}{\ensuremath{\overline{\rm{p}}}\xspace}
\newcommand{\ppr}{\ensuremath{{\rm p}}\xspace}
\newcommand{\pprpm}{\ensuremath{(\pprm) \pprp }\xspace}
\newcommand{\pprs}{\ensuremath{\pprp + \pprm}\xspace}
\newcommand{\pdeu}{\ensuremath{{\rm d}}\xspace}
\newcommand{\pphi}{\ensuremath{\phi}\xspace}
\newcommand{\plam}{\ensuremath{\Lambda}\xspace}
\newcommand{\palam}{\ensuremath{\overline{\Lambda}}\xspace}
\newcommand{\psigp}{\ensuremath{\Sigma^{+}}\xspace}
\newcommand{\rppi}{\ensuremath{\ppr/\ppi}\xspace}
\newcommand{\rphipi}{\ensuremath{\pphi/\ppi}\xspace}

\newcommand{\Mmunu}{\ensuremath{{\rm M}(\mu \nu_{\mu})}\xspace}
\newcommand{\ndfITS}{\ensuremath{N^{\rm hits}_{\rm ITS}}\xspace}
\newcommand{\chitwo}{\ensuremath{\chi^{2}}\xspace}
\newcommand{\chitwoITS}{\ensuremath{\chi^{2}_{\rm ITS}}\xspace}
\newcommand{\chitwoTPC}{\ensuremath{\chi^{2}_{\rm TPC}}\xspace}
\newcommand{\chitwoNDF}{\ensuremath{\chi^{2}/{\rm NDF}}\xspace}
\newcommand{\dca}{\ensuremath{{\rm DCA}}\xspace}
\newcommand{\dcaXY}{\ensuremath{\dca_{\it{xy}}}\xspace}
\newcommand{\sigmadcaXY}{\ensuremath{\sigma_{\dcaXY}}\xspace}
\newcommand{\dcaZ}{\ensuremath{\dca_{\it{z}}}\xspace}
\newcommand{\mom}{\ensuremath{p}\xspace}
\newcommand{\momTRUE}{\ensuremath{\mom^{\rm TRUE}}\xspace}
\newcommand{\momMEAS}{\ensuremath{\mom^{\rm MEAS}}\xspace}
\newcommand{\pt}{\ensuremath{p_{\rm{T}}}\xspace}
\newcommand{\ptTRUE}{\ensuremath{\pt^{\rm TRUE}}\xspace}
\newcommand{\ptMEAS}{\ensuremath{\pt^{\rm MEAS}}\xspace}
\newcommand{\meanpt}{\ensuremath{\langle \pt \rangle}\xspace}
\newcommand{\mt}{\ensuremath{m_{\rm{T}}}\xspace}
\newcommand{\qt}{\ensuremath{q_{\rm{T}}}\xspace}
\newcommand{\cmc}{\ensuremath{\text{cm}/c}\xspace}
\newcommand{\texpi}{\ensuremath{t_{\rm exp}(i)}\xspace}
\newcommand{\texp}{\ensuremath{t_{\rm exp}}\xspace}
\newcommand{\evtime}{\ensuremath{t_{\rm ev}}\xspace}
\newcommand{\ttof}{\ensuremath{t_{\rm TOF}}\xspace}
\newcommand{\Nsigma}{\ensuremath{{\rm N}_{\sigma}}\xspace}
\newcommand{\nsigma}{\ensuremath{\Nsigma}\xspace}
\newcommand{\Nsigmai}{\ensuremath{{\rm N}_{\sigma, i}}\xspace}
\newcommand{\NsigmaTPC}{\ensuremath{{\rm N}_{\sigma}^{\rm TPC}}\xspace}
\newcommand{\NsigmaTOF}{\ensuremath{{\rm N}_{\sigma}^{\rm TOF}}\xspace}
\newcommand{\tofmeas}{\ensuremath{\text{time-of-flight}}\xspace}
\newcommand{\AAaa}{\ensuremath{{\rm{AA}}}\xspace}
\newcommand{\raa}{\ensuremath{R_{\AAaa}}\xspace}
\newcommand{\taa}{\ensuremath{T_{\AAaa}}\xspace}
\newcommand{\ncoll}{\ensuremath{N_{\rm{coll}}}\xspace}
\newcommand{\dedx}{\ensuremath{{\rm d}E/{\rm d}x}\xspace}
\newcommand{\meandedx}{\ensuremath{\langle\dedx\rangle}\xspace}
\newcommand{\dndyV}{\ensuremath{{\rm d}N/{\rm d}y}\xspace}
\newcommand{\Nch}{\ensuremath{{\it N}_{\rm{ch}}}\xspace}
\newcommand{\dNchdeta}{\ensuremath{\rm{d}\Nch/\rm{d}\eta}\xspace}
\newcommand{\avdNchdeta}{\ensuremath{\langle\dNchdeta\rangle}\xspace}
\newcommand{\dndydpt}{\ensuremath{{\rm d}^{2}N/{\rm d}y{\rm d}\pt}\xspace}
\newcommand{\Tch}{\ensuremath{T_{\rm{ch}}}\xspace}
\newcommand{\Tc}{\ensuremath{T_{\rm{c}}}\xspace}
\newcommand{\Tkin}{\ensuremath{T_{\rm{kin}}}\xspace}
\newcommand{\Bt}{\ensuremath{\beta_{\rm{T}}}\xspace}
\newcommand{\avBt}{\ensuremath{\langle\Bt\rangle}\xspace}
\newcommand{\EcrossB}{\ensuremath{E\times B}\xspace}
\newcommand{\centint}[2]{\ensuremath{#1-#2\%}\xspace}
\newcommand{\momint}[2]{\ensuremath{#1-#2~\gevc}\xspace}
\newcommand{\minv}{\ensuremath{M_{\rm{KK}}}\xspace}
\newcommand{\twopi}{\ensuremath{2\pi}\xspace}
\newcommand{\fun}[2]{\ensuremath{#1\left(#2\right)}}

\newcommand{\xexe}{\ensuremath{\text{Xe--Xe}}\xspace}
\newcommand{\pbpb}{\ensuremath{\text{Pb--Pb}}\xspace}
\newcommand{\ppb}{\ensuremath{\text{p--Pb}}\xspace}
\newcommand{\pp}{\ensuremath{\text{pp}}\xspace}
\newcommand{\gevcV}{\ensuremath{{\rm GeV}/c}\xspace}
\newcommand{\mevcV}{\ensuremath{{\rm MeV}/c}\xspace}
\newcommand{\kevcV}{\ensuremath{{\rm keV}/c}\xspace}
\newcommand{\tevcV}{\ensuremath{{\rm TeV}/c}\xspace}
\newcommand{\kevcsq}{\ensuremath{{\kevcV}^{2}}\xspace}
\newcommand{\mevcsq}{\ensuremath{{\mevcV}^{2}}\xspace}
\newcommand{\gevcsq}{\ensuremath{{\gevcV}^{2}}\xspace}
\newcommand{\mev}{\ensuremath{{\rm MeV}}\xspace}

\newcommand{\sF}{\ensuremath{\s~=~5.02}\xspace}
\newcommand{\snn}{\ensuremath{\sqrt{s_{\rm{NN}}}}\xspace}
\newcommand{\snnT}{\ensuremath{\snn~=~2.76}\xspace}
\newcommand{\snnF}{\ensuremath{\snn~=~5.02}\xspace}
\newcommand{\snnXeXe}{\ensuremath{\snn~=~5.44}\xspace~TeV~}

\newcommand{\Refs}[1]{Refs.~\cite{#1}\xspace}
\newcommand{\Tab}[1]{Tab.~\ref{#1}\xspace}
\newcommand{\Eq}[1]{Eq.~\ref{#1}\xspace}
\newcommand{\Figs}[1]{Figs.~\ref{#1}\xspace}
\newcommand{\Fig}[1]{Fig.~\ref{#1}\xspace}
\newcommand{\Sec}[1]{Section~\ref{#1}\xspace}
\newcommand{\Ustat}{Stat. Uncert.\xspace}
\newcommand{\Usyst}{Syst. Uncert.\xspace}
\newcommand{\BR}{B.R.\xspace}

\iffalse
  \newcommand{\comment}[1]{}
  \newcommand{\commentNJ}[1]{}
  \newcommand{\addition}[1]{}
  \newcommand{\todoSS}[1]{}
  \newcommand{\todoFN}[1]{}
  \newcommand{\todoNJ}[1]{}
  \newcommand{\todoPA}[1]{}
\else
  \newcommand{\comment}[1]{\textcolor{red}{\textit{#1}}}
  \newcommand{\commentNJ}[1]{\textcolor{blue}{[NJ: \textit{#1}]}}
  \newcommand{\addition}[1]{\textcolor{blue}{#1}}
  \newcommand{\todoNJ}[1]{\textcolor{red}{To do: Nicol\`o #1}}
  \newcommand{\todoDK}[1]{\textcolor{red}{To do: David K. #1}}
  \newcommand{\todoCDM}[1]{\textcolor{red}{To do: Chiara #1}}
\fi

\newcommand{\GEANTT}{\ensuremath{\text{GEANT3}}\xspace}
\newcommand{\GEANTF}{\ensuremath{\text{GEANT4}}\xspace}
\newcommand{\FLUKA}{\ensuremath{\text{FLUKA}}\xspace}
\newcommand{\HIJING}{\ensuremath{\text{HIJING}}\xspace}

\newcommand{\Raa}{\ensuremath{\rm R_{AA}}\xspace}
\newcommand{\Rpa}{\ensuremath{\rm R_{pA}}\xspace}
\newcommand{\Rppb}{\ensuremath{R_{\rm pPb}}\xspace}

\newcommand{\cmsquared}{\ensuremath{{\rm cm}^{2}}\xspace}
\newcommand{\cme}{\ensuremath{\sqrt{s}}\xspace}
\newcommand{\gev}{\ensuremath{{\rm GeV}/c}\xspace}
\newcommand{\auau}{\ensuremath{\rm Au\!-\!Au}\xspace}
\newcommand{\ptopi}{\ensuremath{{\rm p } / \pi}\xspace}
\newcommand{\ktopi}{\ensuremath{({\rm K}^{+}+{\rm K}^{-}) / (\pi^{+}+\pi^{-})}\xspace}
\newcommand{\twotwo}{\ensuremath{2\rightarrow 2}\xspace}
\newcommand{\ltok}{\ensuremath{({\rm \Lambda}^{0}+\bar {\rm \Lambda}^{0})/(\rm 2 K^{0}_{s} )} \xspace}

\newcommand{\snnt}[1]{\ensuremath{\snn~=~#1~\text{\,TeV}}\xspace}
\newcommand{\snnnotext}[1]{\ensuremath{\snn~=~#1}\xspace}
\newcommand{\sppt}[1]{\ensuremath{\sqrt{s} = #1 \text{\,TeV}}\xspace}
\newcommand{\sppg}[1]{\ensuremath{\sqrt{s} = #1 \text{\,GeV}}\xspace}
\newcommand{\gevc}[1]{\ensuremath{#1\ \text{\,\gevcV}}\xspace}
\newcommand{\mevc}[1]{\ensuremath{#1\ \text{\,\mevcV}}\xspace}

\newcommand{\dndetaold}[1]{\ensuremath{\frac{\text{d}^2N_{#1}}{\text{d}\pt \text{d}\eta}}\xspace}
\newcommand{\dndeta}       {\ensuremath{\mathrm{d}N_\mathrm{ch}/\mathrm{d}\eta}\xspace}
\newcommand{\avdndeta}     {\ensuremath{\langle\dndeta\rangle_{|\eta|< 0.5}}\xspace}
\newcommand{\dndetashort}     {\ensuremath{\langle\dndeta\rangle}\xspace}
\newcommand{\dndy}[1]{\ensuremath{\frac{d^2N_{#1}}{d\pt dy}}\xspace}
\newcommand{\eff}[1]{\ensuremath{\epsilon_{#1}}\xspace}
\newcommand{\bareyield}{\ensuremath{Y}\xspace}
\newcommand{\yield}[1]{\ensuremath{Y_{#1}}\xspace}
\newcommand{\ch}{\ensuremath{\text{ch}}\xspace}
\newcommand{\etalab}{\ensuremath{\eta_{lab}}\xspace}
\newcommand{\etaint}[1]{\ensuremath{|\eta| < #1}\xspace}
\newcommand{\yint}[1]{\ensuremath{|y| < #1}\xspace}

\newcommand{\VZ}{\ensuremath{V^{0}}\xspace}
\newcommand{\VZs}{\ensuremath{V^{0}\text{s}}\xspace}
\newcommand{\RAA}{\ensuremath{R_{\text{AA}}}\xspace}
\newcommand{\MRAA}{\ensuremath{\mathbf{R_{\text{\textbf{AA}}}}}\xspace}
\newcommand{\dpi}{\ensuremath{\Delta_{\pi}}\xspace}
\newcommand{\dkaon}{\ensuremath{\Delta_{K}}\xspace}
\newcommand{\dproton}{\ensuremath{\Delta_{p}}\xspace}
\newcommand{\mdpi}{\ensuremath{\mathbf{\Delta_{\pi}}}\xspace}
\newcommand{\mdkaon}{\ensuremath{\mathbf{\Delta_{K}}}\xspace}
\newcommand{\mdproton}{\ensuremath{\mathbf{\Delta_{p}}}\xspace}
\newcommand{\rpi}{\ensuremath{R_{\pi}}\xspace}
\newcommand{\mathdedx}{\ensuremath{\mathbf{\text{d}E/\text{d}x}}\xspace}
\newcommand{\dEdx}         {\ensuremath{\textrm{d}E/\textrm{d}x}\xspace}
\newcommand{\mdedx}{\ensuremath{\left <\text{d}E/\text{d}x \right>}\xspace}
\newcommand{\mathmdedx}{\ensuremath{\mathbf{\left <\text{d}E/\text{d}x \right>}}\xspace}
\newcommand{\mdedxpi}{\ensuremath{\left <\text{d}E/\text{d}x \right>_{\pi}}\xspace}
\newcommand{\meanp}{\ensuremath{\langle p \rangle}\xspace}
\newcommand{\sdedx}{\ensuremath{\sigma_{\text{d}E/\text{d}x}}\xspace}
\newcommand{\relres}{\ensuremath{\sigma/\left <\text{d}E/\text{d}x \right>}\xspace}
\newcommand{\res}{\ensuremath{\sigma_{\text{d}E/\text{d}x}}\xspace}
\newcommand{\ncl}{\ensuremath{\text{Ncl}}\xspace}
\newcommand{\mncl}{\ensuremath{\langle \text{Ncl} \rangle}\xspace}

\newcommand{\chpi}{\ensuremath{\pi^{+}+\pi^{-}}\xspace}
\newcommand{\chk}{\ensuremath{{\rm K}^{+}+{\rm K}^{-}}\xspace}
\newcommand{\chp}{\ensuremath{{\rm p}+{\rm \bar{p}}}\xspace}

\newcommand{\bg}{\ensuremath{\beta\gamma}\xspace}

\newcommand{\timemeasure}{\ensuremath{{t}\xspace}}
\newcommand{\Otwo}{\ensuremath{{\rm O}^{2}}\xspace}
\newcommand{\LA}{\ensuremath{\Lambda}\xspace}
\newcommand{\AL}{\ensuremath{\bar{\Lambda}}\xspace}
\newcommand{\KOs}{\ensuremath{\rm K^0_S}\xspace}
\newcommand{\dMassG}{\ensuremath{\Delta m_\gamma}\xspace}
\newcommand{\dMassL}{\ensuremath{\Delta m_\Lambda}\xspace}
\newcommand{\dMassAL}{\ensuremath{\Delta m_{\bar{\Lambda}}}\xspace}
\newcommand{\dMassKO}{\ensuremath{\Delta m_{K^0_s}}\xspace}
\newcommand{\TFF}{\ensuremath{\texttt{TFractionFitter}}\xspace}
\newcommand{\RooFit}{\ensuremath{\texttt{RooFit}}\xspace}
\newcommand{\deltaOne}{\ensuremath{{\rm (\deltaPi)_{1}}}\xspace}
\newcommand{\deltaTwo}{\ensuremath{{\rm (\deltaPi)_{2}}}\xspace}
\newcommand{\sigmatoftrk}{\ensuremath{{\rm \sigma_{TOF} \oplus \sigma_{Trk}}}\xspace}
\newcommand{\doubledelta}{\ensuremath{{\rm \Delta \Delta} \timemeasure}\xspace}
\newcommand{\doubledeltaLong}{\ensuremath{\deltaTwo - \deltaOne}\xspace}
\newcommand{\texpPi}{\ensuremath{\texp(\pi)}\xspace}
\newcommand{\texpKa}{\ensuremath{\texp(\rm K)}\xspace}
\newcommand{\texpPr}{\ensuremath{\texp(\rm p)}\xspace}
\newcommand{\deltaPi}{\ensuremath{\timemeasure - \texpPi}\xspace}
\newcommand{\deltaKa}{\ensuremath{\timemeasure - \texpKa}\xspace}
\newcommand{\deltaPr}{\ensuremath{\timemeasure - \texpPr}\xspace}
\newcommand{\evTime}{\ensuremath{\timemeasure_{\rm ev.}}\xspace}
\newcommand{\evTimeTOF}{\ensuremath{\evTime^{\rm TOF}}\xspace}
\newcommand{\evTimeTZAC}{\ensuremath{\evTime^{\rm FT0}}\xspace}
\newcommand{\sigmaTOF}{\ensuremath{ \sigma_{\rm TOF} }\xspace}
\newcommand{\sigmaTOFpid}{\ensuremath{ \sigmaTOF^{\rm PID} }\xspace}
\newcommand{\sigmaTrk}{\ensuremath{ \sigma_{\rm Tracking} }\xspace}
\newcommand{\sigmaRef}{\ensuremath{ \sigma_{\rm Reference} }\xspace}

\newcommand{\deltaGeneral}{\ensuremath{\timemeasure - \texp - \evTime}\xspace}
\newcommand{\deltaPiTOF}{\ensuremath{\timemeasure - \texpPi - \evTimeTOF}\xspace}
\newcommand{\deltaKaTOF}{\ensuremath{\timemeasure - \texpKa - \evTimeTOF}\xspace}
\newcommand{\deltaPrTOF}{\ensuremath{\timemeasure - \texpPr - \evTimeTOF}\xspace}
\newcommand{\deltaPiTZAC}{\ensuremath{\timemeasure - \texpPi - \evTimeTZAC}\xspace}
\newcommand{\deltaKaTZAC}{\ensuremath{\timemeasure - \texpKa - \evTimeTZAC}\xspace}
\newcommand{\deltaPrTZAC}{\ensuremath{\timemeasure - \texpPr - \evTimeTZAC}\xspace}
\newcommand{\tofEvMult}{\ensuremath{\rm TOF\ ev. mult.}\xspace}

\newcommand{\relval}{\texttt{rel\_val}\xspace}
\newcommand{\apassthree}{\texttt{apass3}\xspace}
\newcommand{\RunTwo}{\ensuremath{\rm Run~2}\xspace}
\newcommand{\RunThree}{\ensuremath{\rm Run~3}\xspace}
\newcommand{\trkOne}{\ensuremath{\rm track_{1}}\xspace}
\newcommand{\trkTwo}{\ensuremath{\rm track_{2}}\xspace}

\begin{titlepage}
\PHyear{2026}       
\PHnumber{174}      
\PHdate{11 June}  

\title{Precision mass measurements of multistrange baryons
\\~and their antiparticles}
\ShortTitle{Mass measurements of multi-strange baryons}   

\Collaboration{ALICE Collaboration\thanks{See Appendix~\ref{app:collab} for the list of collaboration members}}
\ShortAuthor{ALICE Collaboration} 

\begin{abstract}
The $\Omega^-$ baryon, composed of three strange quarks (sss), was predicted using the quark model and discovered in 1964; it played a pivotal role in establishing quarks as fundamental constituents of matter. Despite its importance, experimental knowledge about its mass remains limited, with the current world average relying on measurements performed more than four decades ago and lacking robust estimates of systematic uncertainties. 
This is notable given the central role of the 
$\Omega^-$ mass, and alternatively that of the $\Xi^-$\,(dss), in lattice quantum chromodynamics calculations, where it is widely used to set the overall physical scale. Precise scale setting is essential for first-principles studies of quark confinement, chiral symmetry breaking, and stringent tests of the Standard Model.
Here we report high-precision measurements of the masses of the 
$\Omega^-$ and $\Xi^-$ baryons and their respective antiparticles,
determined from invariant-mass reconstruction of their decay products in proton--proton collisions at the Large Hadron Collider.
The analysis exploits the excellent tracking and particle-identification capabilities of the ALICE experiment, 
enabling accurate reconstruction of the displaced decay vertices characteristic of these short-lived particles. 
Each mass is measured with a fractional uncertainty of about 60 parts per million, for example $M_{\omp}=1672.558\,\pm\,0.034\,({\scriptstyle \rm stat.})\,\pm\,0.102\,({\scriptstyle \rm syst.})$~MeV/$c^2$.
The precisely known $\pkazero$ and $\plam$ masses are used for calibration. 
These results establish new precision benchmarks in strange-baryon spectroscopy and enable stringent tests of Charge-Parity-Time invariance in the multistrange-hadron sector.
Our measurement reduces the scale uncertainty in lattice quantum chromodynamics calculations, enabling for instance sub per mille precision for the hadronic vacuum-polarization contribution to the muon anomalous magnetic moment.

\end{abstract}
\end{titlepage}

\setcounter{page}{2} 


Quantum chromodynamics (QCD) is the fundamental theory that describes the strong interaction between quarks and gluons~\cite{pdgParticlePhysics2024}. Prior to the development of QCD, the quark model provided a systematic classification of hadrons and was used to predict the existence of the triply strange $\Oms^{-}$ ($\rm sss$) baryon~\cite{Gell-Mann:1964ewy}. Its discovery~\cite{Barnes:1964pd} in 1964 provided direct confirmation of quarks as the fundamental constituents of matter, establishing the quark model as the organizing principle of hadrons and laying the groundwork for the later development of QCD. However, experimental constraints on the $\Oms^{-}$ mass remain remarkably limited.

The current world average of the mass of the 
$\Oms^-$ baryon is $M_\Omega~=~1672.45\,\pm\,0.29$~MeV/$c^2$~\cite{pdgParticlePhysics2024}, obtained assuming equal masses for particle and antiparticle. The contributing measurements are dominated by statistical uncertainties, with either no systematic uncertainty reported~\cite{Baubillier:1978, Hartouni:1985mn} or only partial estimates available~\cite{Hemingway:1978}, leaving the overall accuracy of the world average difficult to assess.
In comparison, the mass of the doubly-strange $\Xis^-$ ($\rm dss$) baryon has a firmer experimental basis: the world average value is $M_\Xi~=~1321.71\,\pm\,0.07$~MeV/$c^2$~\cite{pdgParticlePhysics2024}. However, the quoted precision is driven by a single measurement, accounting for both statistical and systematic uncertainties~\cite{abdallahMassesLifetimesProduction2006}, and an independent experimental confirmation would be very valuable.

Lattice QCD provides a first-principles framework for describing fundamental phenomena such as the confinement of quarks and gluons into color-singlet hadrons~\cite{Fodor:2012gf,durrInitioDeterminationLight2008a}. It also offers deep insights into the non-trivial structure of the QCD vacuum and the mechanism of spontaneous chiral symmetry breaking, which generates approximately 99\% of the nucleon mass and thus accounts for nearly all the visible mass of the universe. 

In this context, the $\Oms^{-}$ baryon mass, or alternatively that of the $\Xis^{-}$ baryon, is commonly used to set the overall physical scale in lattice QCD studies~\cite{Fodor:2012gf}. 
Such baryons are relatively heavy and the lightest (u,d) quarks are either absent or of minor importance among their valence quarks.
Consequently, the $\Xis^{-}$ and preferentially $\Oms^{-}$ masses can be determined with high precision in lattice QCD calculations. 
Here, the dimensionless lattice spacing $a$ is given a physical unit by comparing the calculated mass (in lattice units) with the experimental one, $a_{\rm physical} = (a\cdot M_{\Omega_{\rm lattice}})/M_{\Omega_{\rm exp}}$. Given this central role of the $\Oms^{-}$ baryon, the scarcity of experimental data is particularly noteworthy.

Eventually, stringent comparisons of particle and antiparticle masses allow direct tests of CPT invariance, a fundamental symmetry of relativistic quantum field theories that asserts that the combined operations of Charge conjugation (C), Parity transformation (P), and Time reversal (T) leave the laws of physics invariant~\cite{Luders:1954zz, streater1989pct}. It implies, for example, that particles and their antiparticles have identical masses, lifetimes and equal but opposite charges and magnetic moments~\cite{PhysRev.106.385}. A violation of CPT invariance would constitute a clear evidence for new physics and a departure from the Standard Model framework.\\
The quark content of the $\Omega^-$ baryon ($\rm sss$)  makes it distinct from lighter baryons composed of up and down quarks. 
Its unique quark composition and higher mass make it a sensitive probe for CPT symmetry tests, 
as potential violations may depend on quark flavour or mass scale~\cite{Kostelecky:1996fk, Kostelecky:2003fs}, 
thereby extending such studies into the multistrange-baryon sector, beyond the well-studied cases of protons (uud) and neutrons (udd).

Unlike protons which can be stored for extended periods of time in Penning traps~\cite{disciaccaDirectMeasurementProton2012},
the multistrange baryons $\Xi$ and $\Omega$ are difficult to produce, and they decay through the weak interaction with lifetimes of a fraction of a nanosecond~\cite{pdgParticlePhysics2024}. As a result, obtaining high-precision measurements of their properties is considerably more challenging. Reducing the uncertainty on the mass values requires large datasets and detectors with excellent tracking and momentum resolution.

The Large Hadron Collider (LHC) at CERN provides 
proton--proton (pp) collisions at the world's highest energies~\cite{evansLHCMachine2008}, in which multistrange baryons and antibaryons are produced in nearly equal quantities, with the antibaryon-to-baryon ratios exceeding 99\% and consistent with unity~\cite{alice2013antibaryonToBaryonRatiosInPP7TeV}. 
The ALICE detector~\cite{alicecollaborationALICEExperimentCERN2008} at the LHC is particularly well suited for studying strange baryons. Within the central-barrel acceptance (pseudorapidity $|\eta| < 0.8$), the experiment offers excellent momentum resolution over a wide momentum range (approximately $1-2$\% for transverse momenta from 1 to 10 \gevcV~\cite{alicecollaborationPerformanceALICEExperiment2014}), enabling high-precision measurements; 
robust particle-identification capabilities, ensuring precise control of the combinatorial background; 
and precise secondary vertex resolution ($\sim$100 $\mu$m for transverse momenta above 1 \gevcV~\cite{brunaVertexReconstructionProtonproton2009}), allowing accurate reconstruction of weak-decay topologies. 

In this work, high-precision measurements of the mass of the $\Omega^{-}$ and $\Xi^{-}$ baryons, along with their antiparticles, are reported using the precisely known masses of the singly-strange hadrons \pkazero and $\Lambda$ as standard candles for calibration.
By exploiting high tracking resolution and large samples of proton--proton collision data, each measurement achieves an unprecedented fractional uncertainty of typically 60 parts per million. 

The analysed data sample consists of proton--proton collisions at a centre-of-mass energy of $\sqrt{s} = 13$~TeV, collected by the ALICE detector at the LHC during the 2016–2018 data-taking period.  The main detectors employed in this analysis are the Inner Tracking System (ITS)~\cite{alicecollaborationAlignmentALICEInner2010} --- made of six concentric layers of silicon pixel, drift and strip detectors --- for reconstructing primary and secondary vertices, and the Time Projection Chamber
(TPC)~\cite{almeALICETPCLarge2010} for providing momentum measurements and identifying charged particles based on their energy
loss in the TPC gas. Both ITS and TPC are embedded within a solenoidal magnet providing a homogeneous magnetic field of $\pm$ 0.5 T, directed in the $z$-direction along the beam axis. 

\begin{figure}[t]
\begin{center}
\includegraphics[width=0.8\linewidth,clip]{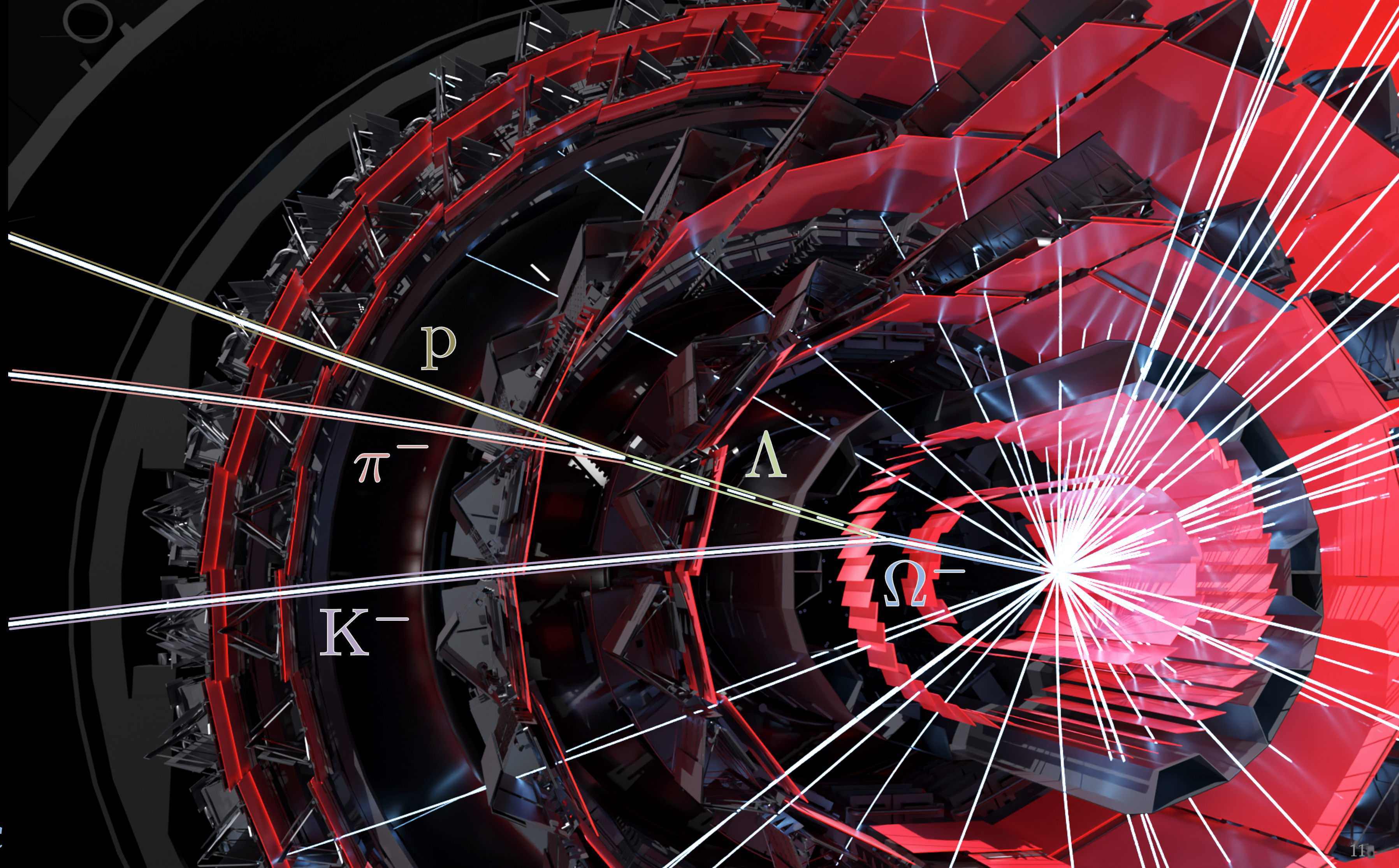}
\end{center}
\label{fig:XiEventDisplay} 
\caption{A typical $\Oms$ decay in the ALICE detector, with the 6 silicon layers of ITS, with the active detector elements in red and support structures in grey. The trajectories of the $\Oms$, $\plam$, kaon, proton and pion trajectories are indicated with borders in light blue, light green (dashed line), light violet, olive-green and red-orange, respectively.}\label{fig:EvtDisplay}
\end{figure}

The multistrange baryons are reconstructed at midrapidity ($0 < y < 0.5$, see Methods) through their decay channels: $\Omega^{-} \rightarrow {\rm K}^{-} \Lambda \rightarrow {\rm K}^{-} \pi^{-}$p (with an integrated branching ratio B.R.~=~43.4\%) and $\Xi^{-} \rightarrow \pi^{-} \Lambda \rightarrow \pi^{-} \pi^{-}$p (B.R.~=~63.9\%)
~\cite{pdgParticlePhysics2024}; further details are provided in Table~\ref{tab:V0CascPDGMass} in the Methods section.
Similarly, the corresponding antibaryons $\overline{\Omega}^{+}$ and $\overline{\Xi}^{+}$ are reconstructed through the charge conjugates along the same decay chains.
Figure~\ref{fig:EvtDisplay} shows an event display with the typical decay of an $\Omega^{-}$ in the ALICE detector. The charged $\Omega$ originates from the proton--proton collision vertex, and propagates up to a few centimetres ($c \tau = 2.461$~cm) before decaying via the weak interaction into a neutral $\Lambda$($\rm uds$) hyperon and a charged kaon. In turn, the $\Lambda$ hyperon ($c \tau = 7.845$~cm) further decays into a proton and a pion, in a topology also known as a $\rm V^{0}$ decay. This decay chain in two steps is usually referred to as a \emph{cascade} decay. The $\Xi$ baryon has an equivalent cascade topology, but with a longer lifetime ($c\tau = 4.91$ cm).

The reconstruction of cascade decay topologies is performed using the trajectories of the decay daughters derived from combined measurements in the ITS and the TPC detectors. 
Tracks with opposite electric charge are first paired to form $\Lambda$ candidates, which are then combined with a charged pion or kaon to reconstruct the full cascade decay. To reduce combinatorial background and enhance signal purity, a set of spatial and kinematic selection criteria is strategically applied, described in the Methods section~\ref{ssec:TopologicalReco}. Additional background reduction and particle identification are achieved by comparing the measured average specific energy loss, $\langle \dEdx \rangle$, with the expected values under various mass hypotheses.

A crucial ingredient for achieving the present measurement precision is the control and correction of reconstruction-related biases. In the standard ALICE tracking framework~\cite{alicecollaborationPerformanceALICEExperiment2014}, tracks are extrapolated to the collision point under some approximations — a uniform material composition and a mass hypothesis not necessarily coinciding with the true particle mass — introducing momentum biases for the decay products of weakly-decaying particles.
For this analysis, a dedicated procedure is employed, which propagates tracks to the secondary vertex, accounting for the actual crossed material and using the appropriate mass hypothesis for each particle species. Validated with Monte Carlo simulations including realistic detector response and tracking performance, this approach significantly reduces momentum biases and is critical for the accuracy of the measurement.

To further control the reconstruction-related biases and ensure a stable measurement, the cascade candidates are reconstructed at intermediate values of transverse momentum (typically $1.4 < p_{\rm T} < 5~\gev$, with the exact range depending on the particle species, see Methods) --- avoiding the low-momentum region dominated by multiple scattering and the high-momentum regime where tracking resolution degrades --- and for small opening angles between the $\Lambda$ and the $\pi^{\pm}$ (or ${\rm K}^{\pm}$). 
These variables are known to correlate with mass shifts arising from tracking resolution and vertex-reconstruction performance. Such selections provide effective and experimentally robust means to suppress these biases. In addition, only candidates whose trajectories and decay products lie entirely within the same longitudinal half of the detector are considered, specifically the positive $z$-side ($z$ > 0), corresponding to positive pseudorapidity of the decay products ($\eta_{\rm dau.} > 0$). This restriction helps to control systematic effects related to detector calibration, and to potential mismatches for particle daughters propagating from one half of the detector to the other.

The mass of each cascade candidate is reconstructed from the four-momenta of its decay daughters, exploiting energy-momentum conservation in the decay process. For each cascade candidate, the invariant mass $\frac{1}{c^{2}}\sqrt{ \left(\sum_{i} E_{i}\right)^2 - \left(\sum_{i} {\bf p}_{i} \right)^{2} c^{2}}$, with $E_{i}$ and ${\bf p}_{i}$ being the energy and momentum vector of the $i$-th decay daughter respectively, is calculated by assigning the daughter masses according to either the $\Xi$ or $\Omega$ decay hypothesis. The masses of multistrange baryons are extracted via a fit of the obtained invariant-mass distributions by the sum of two functions: one for the signal, one for the background. 
The signal peak comprises the contribution of candidates reconstructed with different momentum resolutions, and is modelled by a pseudo-Gaussian function defined as a sum of three Gaussians of different widths but of  a common mean.
The background, originating mainly from random combinations of unrelated tracks, is modelled with an exponential function. The measured mass $\mu$ corresponds to the centre of the invariant-mass peak, given by the mean of the pseudo-Gaussian function. The width, denoted as $\sigma$, provides an estimate of the experimental invariant-mass resolution.

\begin{figure}[t]
\subfigure[]{
    \includegraphics[width=0.48\linewidth,clip]{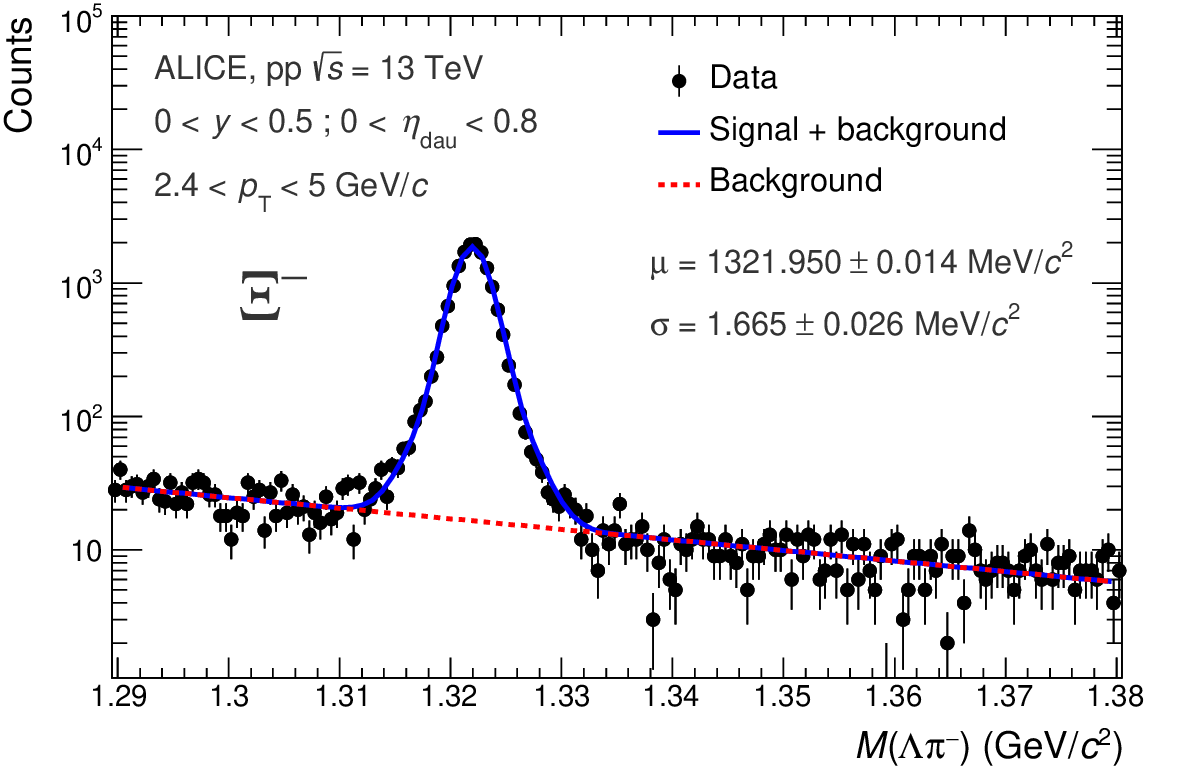}
	\label{fig:XiMinus_TripleGaussian}
}
\subfigure[]{
	\includegraphics[width=0.48\linewidth,clip]{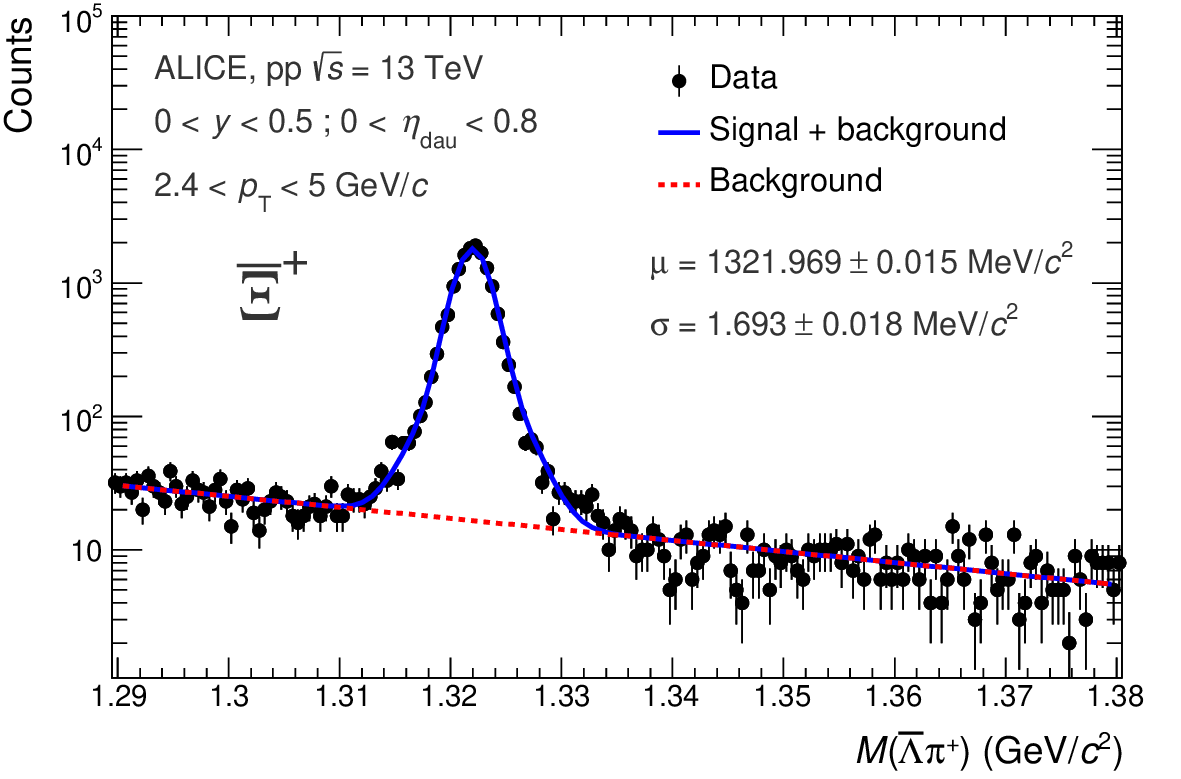}
	\label{fig:XiPlus_TripleGaussian}
} 
\subfigure[]{
	\includegraphics[width=0.48\linewidth,clip]{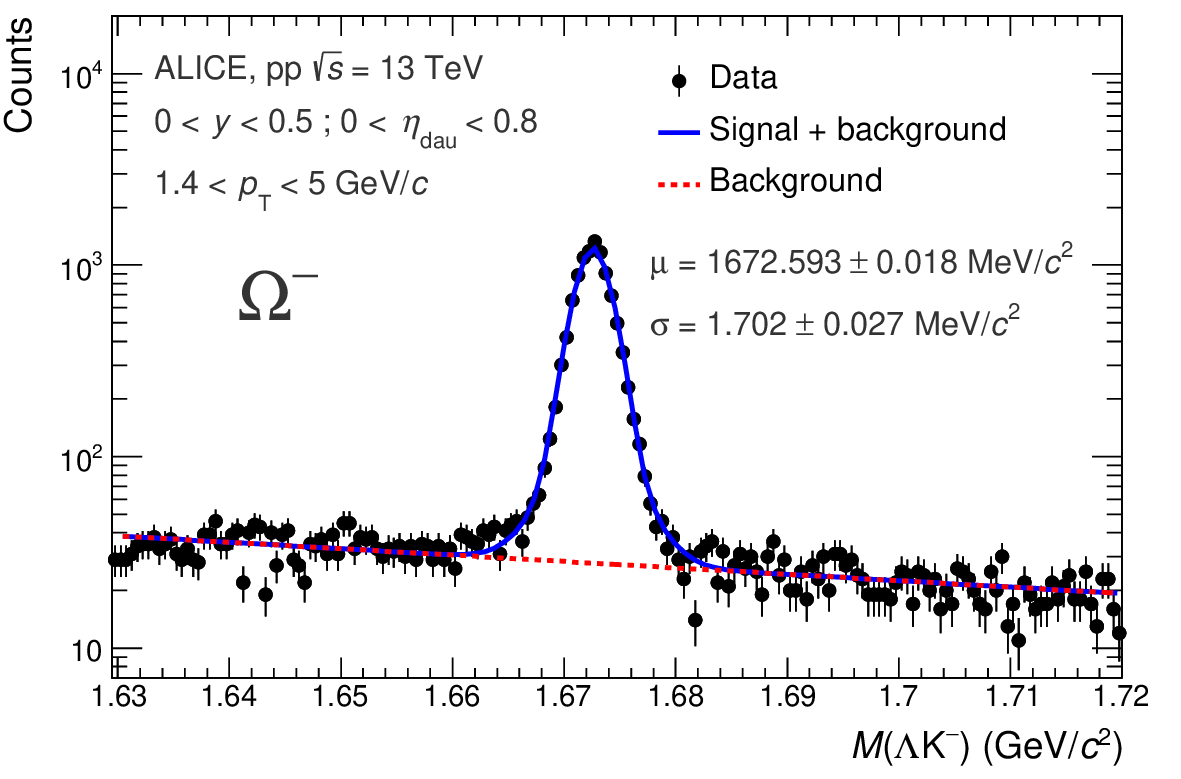}
	\label{fig:OmegaMinus_TripleGaussian}
} 
\subfigure[]{
	\includegraphics[width=0.48\linewidth,clip]{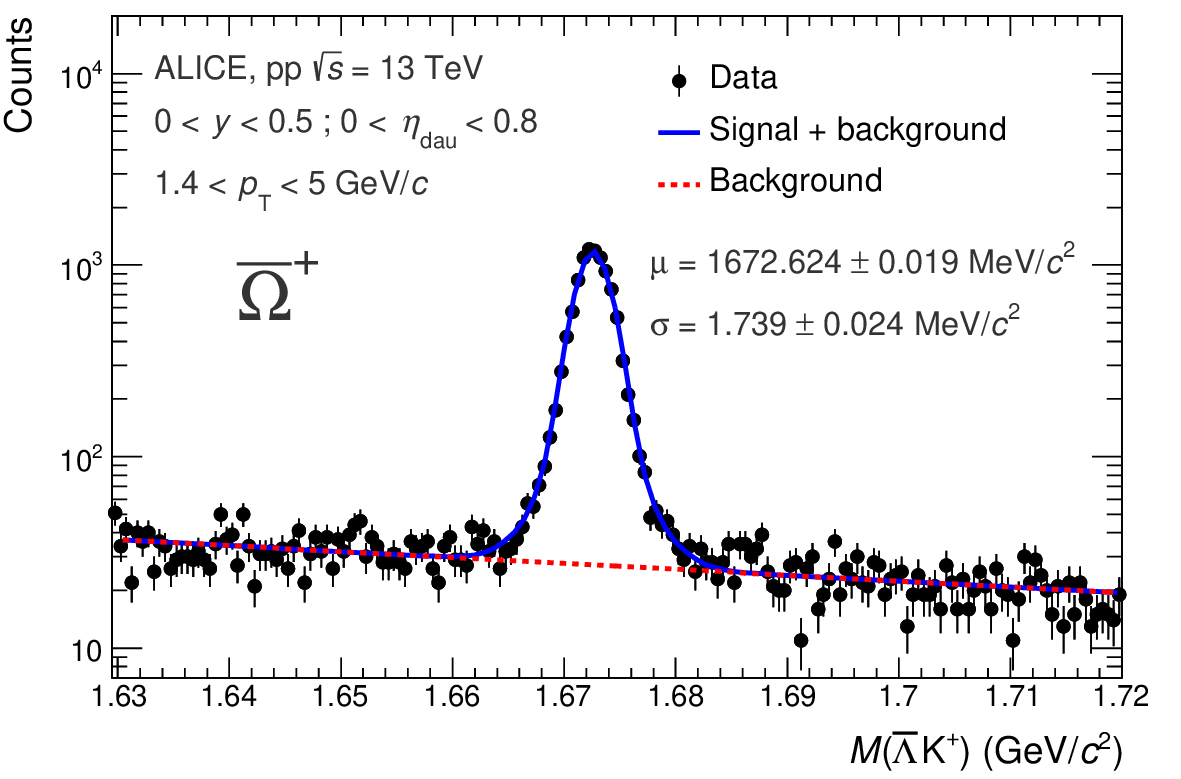}
	\label{fig:OmegaPlus_TripleGaussian}
} 
\caption{Examples of invariant-mass distributions of the $\Xi^{-}$ (\ref{fig:XiMinus_TripleGaussian}), $\overline{\Xi}^{+}$ (\ref{fig:XiPlus_TripleGaussian}), $\Omega^{-}$ (\ref{fig:OmegaMinus_TripleGaussian}) and $\overline{\Omega}^{+}$ (\ref{fig:OmegaPlus_TripleGaussian}). The measured mass $\mu$ and mass resolution $\sigma$ from the fit, with their associated statistical uncertainties, are displayed.}\label{fig:InvMass}
\end{figure}

Examples of invariant-mass distributions of the $\plam \ppim$, $\palam \ppip$, $\plam \pkam$, $\palam \pkap$ pairs obtained in pp collisions at \sppt{13} are shown in Fig.~\ref{fig:InvMass}. The clear peaks corresponding to the \Oms and \Xis baryons demonstrate the excellent particle-identification capabilities of ALICE. The raw \Oms and \Xis yields are greater by factors of one hundred and six, respectively, compared to the samples analysed in previous measurements, while keeping an extraordinarily low level of background, with misidentification contamination of about 9\% and 4\% respectively. Based on the mass parameters extracted from the invariant-mass peaks, the relative mass difference between particle and antiparticle is evaluated as $\Delta M / M = \left( \mu_{\rm part.} - \mu_{\overline{\rm part.}} \right) / \mu_{\rm avg.}$. The corresponding statistical uncertainty is obtained assuming there is no correlation between the particle and antiparticle measurements.

To correct for biases related to data processing, analysis, or fitting procedure, the analysis is applied to Monte Carlo (MC) simulations in which the input masses are set to the PDG values by construction. Such detailed simulations must be modelled as closely as possible on the behaviour and response of the real ALICE apparatus. 
Measured masses in experimental data are corrected for the mass offsets observed in simulations --- defined as the shift between the MC-reconstructed mass and the MC-input mass arising from the reconstruction and selection procedures --- which range from a few tens to a few hundreds of \kevcsq depending on the particle species and the choice of fit function.

The total systematic uncertainties associated with our measurements vary between 78 and 102 \kevcsq depending on the particle species. The dominant contributions originate from the candidate selections, the detector calibration, the finite accuracy of the magnetic-field map and the limited knowledge of the material budget. For mass differences, most systematic effects cancel out due to their symmetric impact on particles and antiparticles. In particular, uncertainties related to the magnetic field map and the material budget are largely correlated between particles and antiparticles and thus cancel to first order. However, residual systematic uncertainties remain, primarily due to differences in the detector response to oppositely charged particles and slight asymmetries in the candidate selection efficiency. Details about the different sources of systematic biases and their evaluation are described in Methods section ~\ref{ssec:SystUncert}.

The measurement accuracy was verified against standard candles, using samples of singly-strange hadrons, exploiting the following decays $\pkazero \rightarrow \pi^{+}\pi^{-}$, $\plam \rightarrow {\rm p} \pi^{-}$ and $\palam \rightarrow \overline{\rm p} \pi^{+}$. 
The masses of \pkazero and \plam are known to great precision, with uncertainties on the world averages down to about 10 \kevcsq: $M_{\pkazero} = 497.611 \pm~0.013$ \mevcsq 
and  $M_{\plam} = 1115.683 \pm 0.006$ \mevcsq~\cite{pdgParticlePhysics2024}; further details can be found in Table~\ref{tab:V0CascPDGMass} in the Methods section. 
In this work, the mass and its uncertainties were determined using the same strategy as for the cascades. 
These three masses and the $\plam-\palam$ mass difference differ by less than $1.5\sigma$ from the respective world averages. The impact of the residual offsets on the cascade masses has been studied and is found to be at the level of 0.05--0.06~\mevcsq, well within the systematic uncertainties of the present measurement, providing a solid anchor for the validity of the cascade results. More details are provided in Methods section~\ref{ssec:SystUncert}.

\begin{figure}[!bp]
\subfigure[]{
    \includegraphics[width=0.48\linewidth,clip]{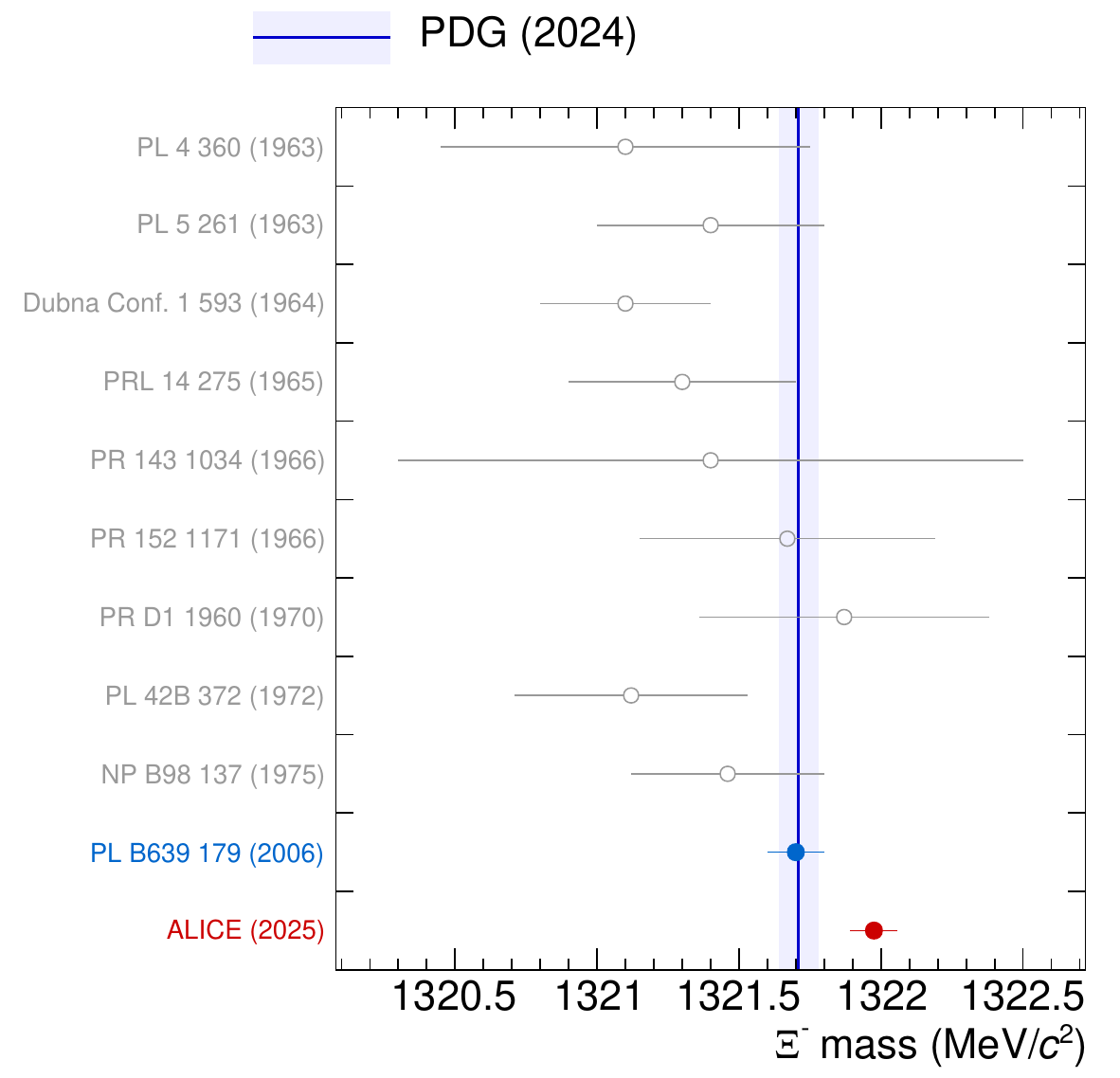}
	\label{fig:FinalMassXiMinus}
}
\subfigure[]{
	\includegraphics[width=0.48\linewidth,clip]{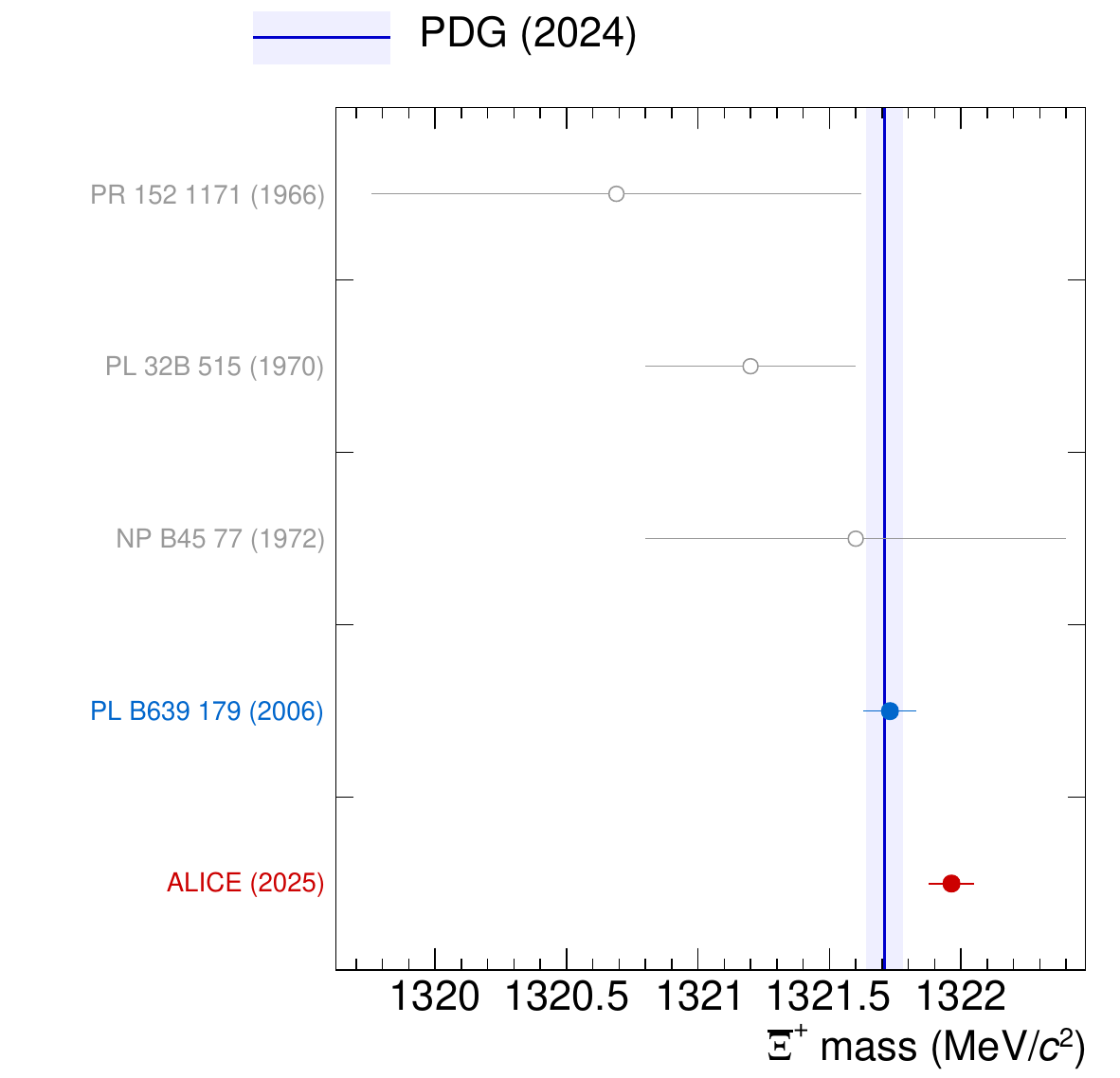}
	\label{fig:FinalMassXiPlus}
} 
\subfigure[]{
	\includegraphics[width=0.48\linewidth,clip]{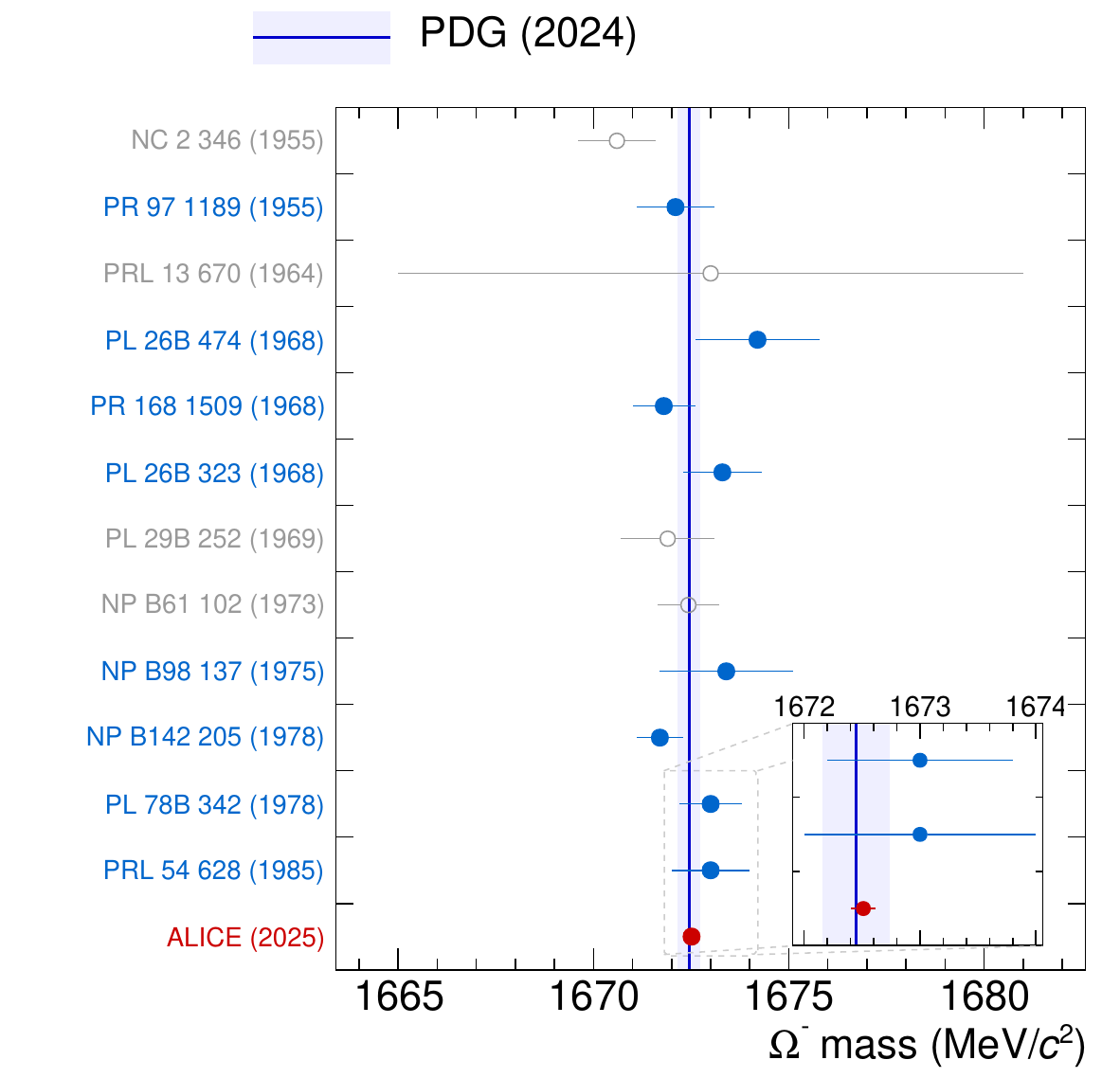}
	\label{fig:FinalMassOmegaMinus}
} 
\subfigure[]{
	\includegraphics[width=0.48\linewidth,clip]{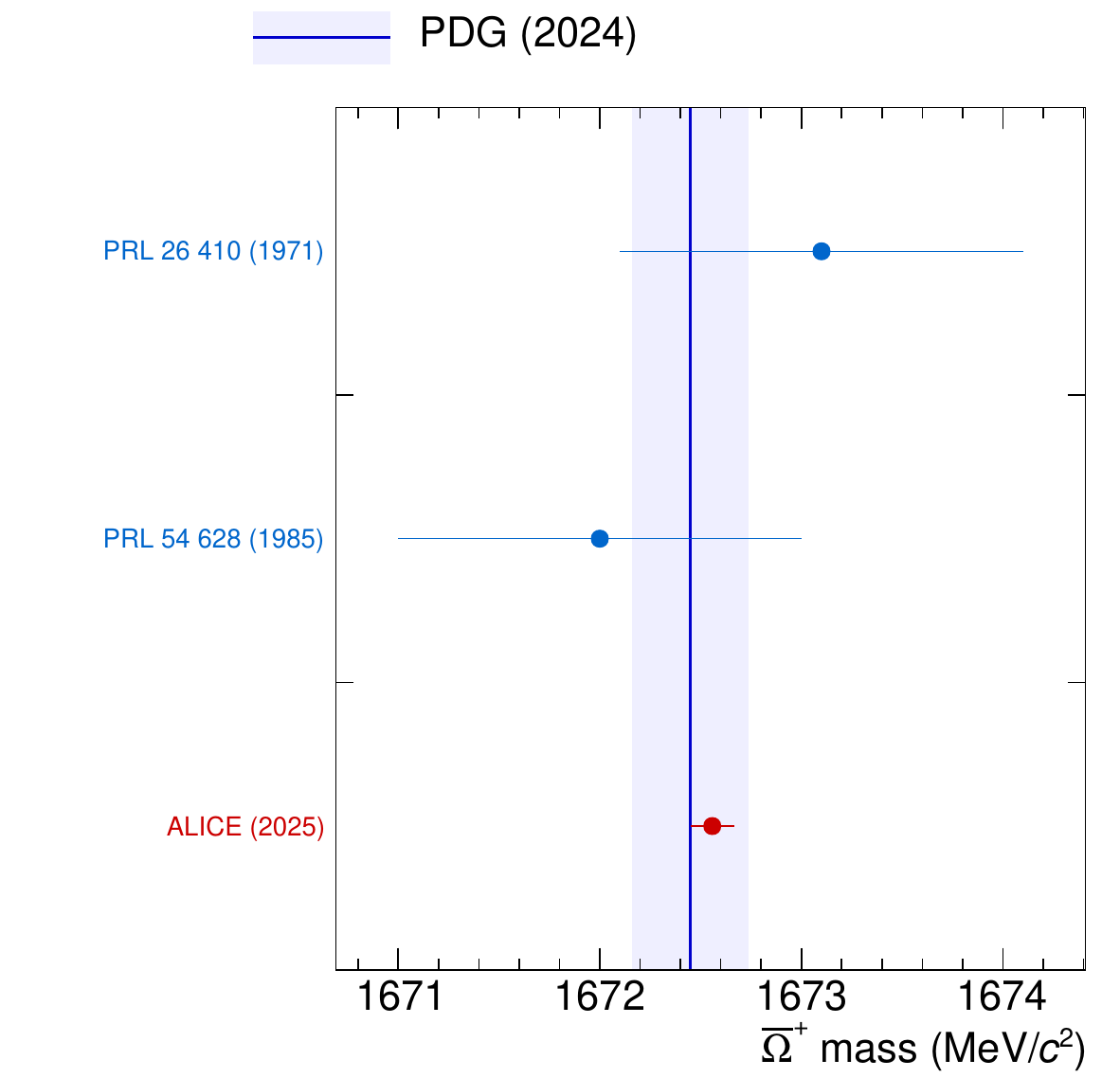}
	\label{fig:FinalMassOmegaPlus}
} 
\caption{Measurements of the mass of $\Xi^{-}$, $\overline{\Xi}^{+}$, $\Omega^{-}$, $\overline{\Omega}^{+}$. The present results are shown in red. The vertical lines and bands represent the mass values currently tabulated by the PDG~\cite{pdgParticlePhysics2024}; The full blue markers represent the measurements on which such PDG average values rely, by contrast to the open grey markers standing for measurements that were discarded.
Note that for the present averages, the PDG assumes \emph{a priori} a strict equality of mass between particle and antiparticle, it implies that the average PDG mass for $\Xi$ stems indifferently from $\Xi^{-}$ and $\overline{\Xi}^{+}$ input. The same situation is met for the PDG $\Omega$ mass with $\Omega^{-}$ and $\overline{\Omega}^{+}$ input.
(See Ref.~\cite{pdgParticlePhysics2024} and reference therein for accessing the individual measurements.)
The horizontal lines correspond to the total uncertainties, obtained by summing the statistical and systematic uncertainties in quadrature.}\label{fig:FinalMass}
\end{figure}

The final results taking into account the statistical and systematic uncertainties are summarised in \mbox{Table}~\ref{tab:FinalResultsCPT}, and a comparison between our mass measurements and previous ones is shown in Fig.~\ref{fig:FinalMass}. The present measurements rely on a sample much larger than those quoted in previous measurements~\cite{pdgParticlePhysics2024}, thanks to the abundant production and detection of multistrange baryons, and thus are no longer dominated by statistical uncertainties. 
The last measurements typically date from experimental campaigns in the early 1990's (DELPHI detector at LEP1, Large Electron-Positron collider, published in 2006, for \Xis~\cite{abdallahMassesLifetimesProduction2006}) 
or even before, using neutral kaon beams (1985 for the most recent \Oms masses published~\cite{Hartouni:1985mn}), where the limited samples available at the time resulted in measurements dominated by statistical uncertainties.\\

\begin{table}[h]
    \caption{Left: Final measured masses and relative mass differences for \Xis and \Oms, with their associated statistical and systematic uncertainties. Right: Previous most precise measurements of the mass and relative mass difference for the \Xis~\cite{abdallahMassesLifetimesProduction2006} and \Oms~\cite{Hartouni:1985mn,Hemingway:1978,chanMeasurementPropertiesOmega1998}, with their total, statistical and systematic uncertainties. Whenever separate values are not quoted in the article, the total uncertainty is indicated.}\label{tab:FinalResultsCPT}
    
    \begin{tabular}{ccccc|cccc}

    \noalign{\smallskip}\hline \noalign{\smallskip}
    \bf Particle & \bf Measured & \multicolumn{3}{c|}{\bf Uncertainty} & \bf Previous & \multicolumn{3}{c}{\bf Uncertainty}\\
    & \bf mass & \bf tot. & \bf stat. & \bf syst. & \bf measured mass & \bf tot. & \bf stat. & \bf syst.\\
    & (\mevcsq) & \multicolumn{3}{c}{(\mevcsq)} & (\mevcsq) & \multicolumn{3}{c}{(\mevcsq)} \\
    \noalign{\smallskip}\hline \noalign{\smallskip}
    \xim & 1321.975 & 0.083 & 0.026 & 0.078 & 1321.70 & 0.10 & 0.08 & 0.05 \\
	\xip & 1321.964 & 0.087 & 0.024 & 0.083 & 1321.73 & 0.10 & 0.08 & 0.05 \\
    \noalign{\smallskip}\hline \noalign{\smallskip}
    \omm & 1672.511 & 0.108 & 0.033 & 0.102 & 1671.7~~ & 0.6~~ & 0.5~~ & 0.3~~ \\ 
    \omp & 1672.558 & 0.108 & 0.034 & 0.102 & 1672~~~~~ & \multicolumn{3}{l}{1} \\ 
	\noalign{\smallskip}\hline \noalign{\smallskip}
	\bf Particle & \bf Measured relative & \multicolumn{3}{c|}{\bf Total} & \bf Previous relative & \multicolumn{3}{c}{\bf Total}\\
    & \bf mass difference & \multicolumn{3}{c|}{\bf uncertainty} & \bf mass difference & \multicolumn{3}{c}{\bf uncertainty} \\
    & ($\times 10^{-5}$) & \multicolumn{3}{c|}{($\times 10^{-5}$)} & ($\times 10^{-5}$) & \multicolumn{3}{c}{($\times 10^{-5}$)}\\
    \noalign{\smallskip}\hline \noalign{\smallskip}
    \Xis & ~~~1.45 & \multicolumn{3}{c|}{6.25} & $-$2.5~~ & \multicolumn{3}{c}{8.7~~~} \\
    \noalign{\smallskip}\hline \noalign{\smallskip}
    \Oms & $-$3.28 & \multicolumn{3}{c|}{4.47} & $-$1.44 & \multicolumn{3}{c}{7.98}\\ 
	\noalign{\smallskip}\hline \noalign{\smallskip}
    \end{tabular}
\end{table}

For both $\Xi$ and $\Omega$ baryons, the measurements reported here are, to date, the most precise obtained by a single experiment.
The precision has been improved by 15\% for the $\Xi$, by factors six and ten for the \omm and \omp, respectively, compared to previous single-experiment measurements. The \omp and \omm mass values are three times more precise than, and consistent with the world average reported by the PDG.
The measured $\Xi$ mass consistently deviates by about 2.2 standard deviations from the world average that is solely determined by the DELPHI measurement~\cite{abdallahMassesLifetimesProduction2006}. 
Together with the excellent agreement observed for the precisely known masses of the singly strange hadrons \pkazero and $\Lambda$, this underscores the necessity of renewed experimental efforts to refine baryon-mass determinations in the strange sector.

Our results for mass differences improve the world average precision by 40\% in the case of the $\Xi$ and by almost a factor of two for the $\Omega$. The differences between particle and antiparticle masses are consistent with zero, in agreement with CPT symmetry, further supporting its validity in the multistrange-baryon sector.

The improved experimental accuracy of the $\Omega$ mass has important implications for precision observables, including for instance the theoretical determination of the muon anomalous magnetic moment $g_{\mu}$.
For two decades, high-precision measurements of $g_{\mu}$ have shown tension with Standard Model predictions, raising the possibility of yet undiscovered forces or elementary particles~\cite{Aliberti:2025beg}. This physical quantity is first and foremost dominated by quantum electrodynamics (QED) considerations but its dominant theoretical uncertainty stems from QCD: it arises from the hadronic vacuum polarization (HVP), in which the photon fluctuates into a virtual quark–antiquark pair, modifying its propagation. This contribution can be computed from first principles using lattice quantum chromodynamics. In such calculations, the HVP is approximately proportional to the square of the lattice spacing, so uncertainties in the scale directly propagate into the predicted value.
As lattice results approach sub‑percent precision~\cite{Boccaletti:2024guq, BMW:2026qvl}, our measurement reduces the uncertainty
in lattice quantum chromodynamics arising from the $\Oms^-$ mass to well below 0.1\%, leaving the remaining uncertainty dominated by other sources.

The recent upgrades of the ALICE apparatus~\cite{alicecollaborationALICEUpgradesLHC2023}, which significantly enhance detection accuracy and collision rate capabilities, will enable precision measurements at the keV scale. This advancement paves the way for fundamental CPT tests in the charmed and beauty baryon sectors in the near future.


\newenvironment{acknowledgement}{\relax}{\relax}
\begin{acknowledgement}
\section*{Acknowledgements}

The ALICE Collaboration would like to thank all its engineers and technicians for their invaluable contributions to the construction of the experiment and the CERN accelerator teams for the outstanding performance of the LHC complex.
The ALICE Collaboration gratefully acknowledges the resources and support provided by all Grid centres and the Worldwide LHC Computing Grid (WLCG) collaboration.
The ALICE Collaboration acknowledges the following funding agencies for their support in building and running the ALICE detector:
A. I. Alikhanyan National Science Laboratory (Yerevan Physics Institute) Foundation (ANSL), State Committee of Science and World Federation of Scientists (WFS), Armenia;
Austrian Academy of Sciences, Austrian Science Fund (FWF): [M 2467-N36] and Nationalstiftung f\"{u}r Forschung, Technologie und Entwicklung, Austria;
Ministry of Communications and High Technologies, National Nuclear Research Center, Azerbaijan;
Rede Nacional de Física de Altas Energias (Renafae), Financiadora de Estudos e Projetos (Finep), Funda\c{c}\~{a}o de Amparo \`{a} Pesquisa do Estado de S\~{a}o Paulo (FAPESP) and The Sao Paulo Research Foundation  (FAPESP), Brazil;
Bulgarian Ministry of Education and Science, within the National Roadmap for Research Infrastructures 2020-2027 (object CERN), Bulgaria;
Ministry of Education of China (MOEC) , Ministry of Science \& Technology of China (MSTC) and National Natural Science Foundation of China (NSFC), China;
Ministry of Science and Education and Croatian Science Foundation, Croatia;
Centro de Aplicaciones Tecnol\'{o}gicas y Desarrollo Nuclear (CEADEN), Cubaenerg\'{\i}a, Cuba;
Ministry of Education, Youth and Sports of the Czech Republic, Czech Republic;
The Danish Council for Independent Research | Natural Sciences, the VILLUM FONDEN and Danish National Research Foundation (DNRF), Denmark;
Helsinki Institute of Physics (HIP), Finland;
Commissariat \`{a} l'Energie Atomique (CEA) and Institut National de Physique Nucl\'{e}aire et de Physique des Particules (IN2P3) and Centre National de la Recherche Scientifique (CNRS), France;
Bundesministerium f\"{u}r Forschung, Technologie und Raumfahrt (BMFTR) and GSI Helmholtzzentrum f\"{u}r Schwerionenforschung GmbH, Germany;
National Research, Development and Innovation Office, Hungary;
Department of Atomic Energy Government of India (DAE), Department of Science and Technology, Government of India (DST), University Grants Commission, Government of India (UGC) and Council of Scientific and Industrial Research (CSIR), India;
National Research and Innovation Agency - BRIN, Indonesia;
Istituto Nazionale di Fisica Nucleare (INFN), Italy;
Japanese Ministry of Education, Culture, Sports, Science and Technology (MEXT) and Japan Society for the Promotion of Science (JSPS) KAKENHI, Japan;
Consejo Nacional de Ciencia (CONACYT) y Tecnolog\'{i}a, through Fondo de Cooperaci\'{o}n Internacional en Ciencia y Tecnolog\'{i}a (FONCICYT) and Direcci\'{o}n General de Asuntos del Personal Academico (DGAPA), Mexico;
Nederlandse Organisatie voor Wetenschappelijk Onderzoek (NWO), Netherlands;
The Research Council of Norway, Norway;
Pontificia Universidad Cat\'{o}lica del Per\'{u}, Peru;
Ministry of Science and Higher Education, National Science Centre and WUT ID-UB, Poland;
Korea Institute of Science and Technology Information and National Research Foundation of Korea (NRF), Republic of Korea;
Ministry of Education and Scientific Research, Institute of Atomic Physics, Ministry of Research and Innovation and Institute of Atomic Physics and Universitatea Nationala de Stiinta si Tehnologie Politehnica Bucuresti, Romania;
Ministerstvo skolstva, vyskumu, vyvoja a mladeze SR, Slovakia;
National Research Foundation of South Africa, South Africa;
Swedish Research Council (VR) and Knut \& Alice Wallenberg Foundation (KAW), Sweden;
European Organization for Nuclear Research, Switzerland;
Suranaree University of Technology (SUT), National Science and Technology Development Agency (NSTDA) and National Science, Research and Innovation Fund (NSRF via PMU-B B05F650021), Thailand;
Turkish Energy, Nuclear and Mineral Research Agency (TENMAK), Turkey;
National Academy of  Sciences of Ukraine, Ukraine;
Science and Technology Facilities Council (STFC), United Kingdom;
National Science Foundation of the United States of America (NSF) and United States Department of Energy, Office of Nuclear Physics (DOE NP), United States of America.
In addition, individual groups or members have received support from:
Czech Science Foundation (grant no. 23-07499S), Czech Republic;
FORTE project, reg.\ no.\ CZ.02.01.01/00/22\_008/0004632, Czech Republic, co-funded by the European Union, Czech Republic;
European Research Council (grant nos. 101220549, 950692), European Union;
Deutsche Forschungs Gemeinschaft (DFG, German Research Foundation) ``Neutrinos and Dark Matter in Astro- and Particle Physics'' (grant no. SFB 1258), Germany;
CONVECS project, CUP C97H23001700002 FESR 2021-2027 program, Italy.

\end{acknowledgement}

\bibliographystyle{utphys}   
\bibliography{aExported_Items}

\newpage
\appendix
\section{Methods}\label{sec:methods}

\begin{table}[h!]
    \caption{A few characteristics, as of 2024, of the \plam, \Xis, \Oms hyperons and the \pkazero meson: quark content, mass, relative mass difference values with their associated uncertainties, dominant decay channel as well as the \mbox{corresponding} branching ratio~\cite{pdgParticlePhysics2024}. The quoted masses correspond to the world averages of the particle and antiparticle masses combined, as tabulated by the PDG.}\label{tab:V0CascPDGMass}
    \begin{adjustwidth}{-0.7cm}{0cm}
    \begin{tabular}{lccccr}
    \noalign{\smallskip}\hline\noalign{\smallskip}
	\multirow{2}{*}{Particle} & Quark & Mass & Relative & Dominant & \multirow{2}{*}{B.R.}\\	
     & content & (\mevcsq) & mass difference & decay channel & \\	
    \noalign{\smallskip}\hline \noalign{\smallskip}
    	
	\pkazero & $\frac{1}{\sqrt{2}}~(\rm d \bar{s} + \bar{d}s)$ & $497.611 \pm 0.013$ & $< 6 \times 10^{-19}$ & \ppip \ppim & 69.2 $\pm$ 0.05~\%\\
	
    \noalign{\smallskip}\hline \noalign{\smallskip}
    
    \plam (\palam) & $\rm u d s$ ($\rm \bar{u}\bar{d}\bar{s}$) & $1115.683 \pm 0.006$ & $\left(-0.1 \pm 1.1\right) \times 10^{-5}$ & \pprp \ppim (\pprm \ppip) & 64.1 $\pm$ 0.5~\% \\
    
    \noalign{\smallskip}\hline \noalign{\smallskip}    
    
    \xim (\xip) & $\rm dss$ ($\rm \bar{d}\bar{s}\bar{s}$) & $1321.71 \pm 0.07$ & $\left(-2.5 \pm 8.7\right) \times 10^{-5}$ & \plam \ppim (\palam \ppip) & 99.887 $\pm$ 0.035~\% \\	
    \noalign{\smallskip}\hline \noalign{\smallskip}
    
	\omm (\omp) & $\rm sss$ ($\rm \bar{s}\bar{s}\bar{s}$) & $1672.45 \pm 0.29$ & $\left(-1.44 \pm 7.98\right) \times 10^{-5}$ & \plam \pkam (\palam \pkap) & 67.7 $\pm$ 0.7~\%\\    
    \noalign{\smallskip}\hline\noalign{\smallskip}
    \end{tabular}
    \end{adjustwidth}
\end{table}

\subsection{Event samples}\label{ssec:EventSamples}

Proton-proton collisions at $\sqrt{s} = 13$~TeV were recorded by the ALICE experiment at the LHC during 2016, 2017 and 2018. The main detectors employed in this analysis are the Inner Tracking System (ITS)~\cite{alicecollaborationAlignmentALICEInner2010}, the Time Projection Chamber (TPC)~\cite{almeALICETPCLarge2010}, the V0~\cite{alicecollaborationPerformanceALICEVZERO2013} and the Time-Of-Flight (TOF)~\cite{akindinovPerformanceALICETimeOfFlight2013}. 
They are embedded within a solenoidal magnet providing a homogeneous magnetic field of $\pm$ 0.5 T, directed in the $z$-direction along the beam axis. The ITS is the closest detection system to the collision point; it is made of six concentric layers of silicon pixel, drift and strip detectors, spanning radii from 3.9 to 43 cm. 
It allows for reconstructing primary vertices of pp collisions as well as secondary vertices from decays of weakly-decaying particles with a precision better than 100 $\mu$m. Surrounding the ITS, the TPC is a cylindrical gaseous detector with an inner radius of 85 cm, an outer radius of 250 cm, and a total length of 500 cm along the beam axis. 
Its active volume is divided into two halves by a central electrode and bounded at both ends by segmented endplates, each composed of 18 trapezoidal sectors equipped with multi-wire proportional chambers. 
The TOF detector envelops the TPC, and is commonly used to identify particles based on their time of flight and plays a crucial role in suppressing out-of-bunch pile-up.
The ITS, TPC and TOF provide full azimuthal coverage and a polar acceptance of about $\pm 45\,^{\textrm{o}}$ with respect to the direction perpendicular to the beam axis. While the ITS yields measurements of up to 6 space points along the trajectory of a charged particle, the TPC yields up to 159 space points. These space points are fitted using a helical track model, which provides the momentum of a charged particle. 
Additionally, the TPC enables particle identification through measurements of the specific energy loss in the gas, \dEdx, which, in combination with the measured momentum, allows long-lived charged hadrons such as pions, kaons and protons to be distinguished from each other.
The V0 system comprises two plastic-scintillator arrays located at forward ($2.8 < \eta < 5.1$) and backward ($-3.7 < \eta < -1.7$) pseudorapidities, with the pseudorapidity defined as $\eta = -\ln[\tan(\theta/2)]$, $\theta$ being the polar angle of the particle with respect to the beam axis. It is primarily used for event triggering. The minimum-bias trigger employed in this analysis requires a coincidence between signals from the two V0 scintillator arrays (V0A and V0C), corresponding to at least one hit in each array. This trigger is thus equivalent to asking for at least two charged particles separated by at least 4.5 units of pseudorapidity. Moreover, the V0 detectors are used to estimate the multiplicity of charged particles by measuring the energy deposited in the scintillators.

To guarantee uniform reconstruction efficiency, the reconstructed primary vertex is required to lie within $\pm$10 cm from the nominal interaction point along the beam axis. Recorded events in which more than one primary vertex is reconstructed, taking into account the position resolution, are discarded to reduce the contamination from pileup (multiple pp collisions in the same bunch crossing).
Beam-induced background and residual pileup events are removed offline using the correlation between the number of track segments found in the two innermost ITS silicon pixel layers and the number of clusters in those layers. In total, the analysed data sample comprises about 2.6 billion minimum-bias pp events at $\sqrt{s} = 13$~TeV.

\subsection{Topological reconstruction}\label{ssec:TopologicalReco}

Direct tracking and identification of the charged \Xis and \Oms hadrons is not possible with the ALICE detector setup used to collect these data as they decay within a few centimetres via the weak interaction into an electrically neutral \plam and a charged particle, the latter also known as the \textit{bachelor}. This initial decay is later followed by the decay of the \plam into a pair of oppositely-charged particles, forming a V-shape topology and hence named a $\rm V^{0}$ particle. See Table~\ref{tab:V0CascPDGMass} for the respective characteristics of implied particles.

\begin{figure}[!h]
\centering
    \includegraphics[angle=0, width=0.4\textheight,clip]{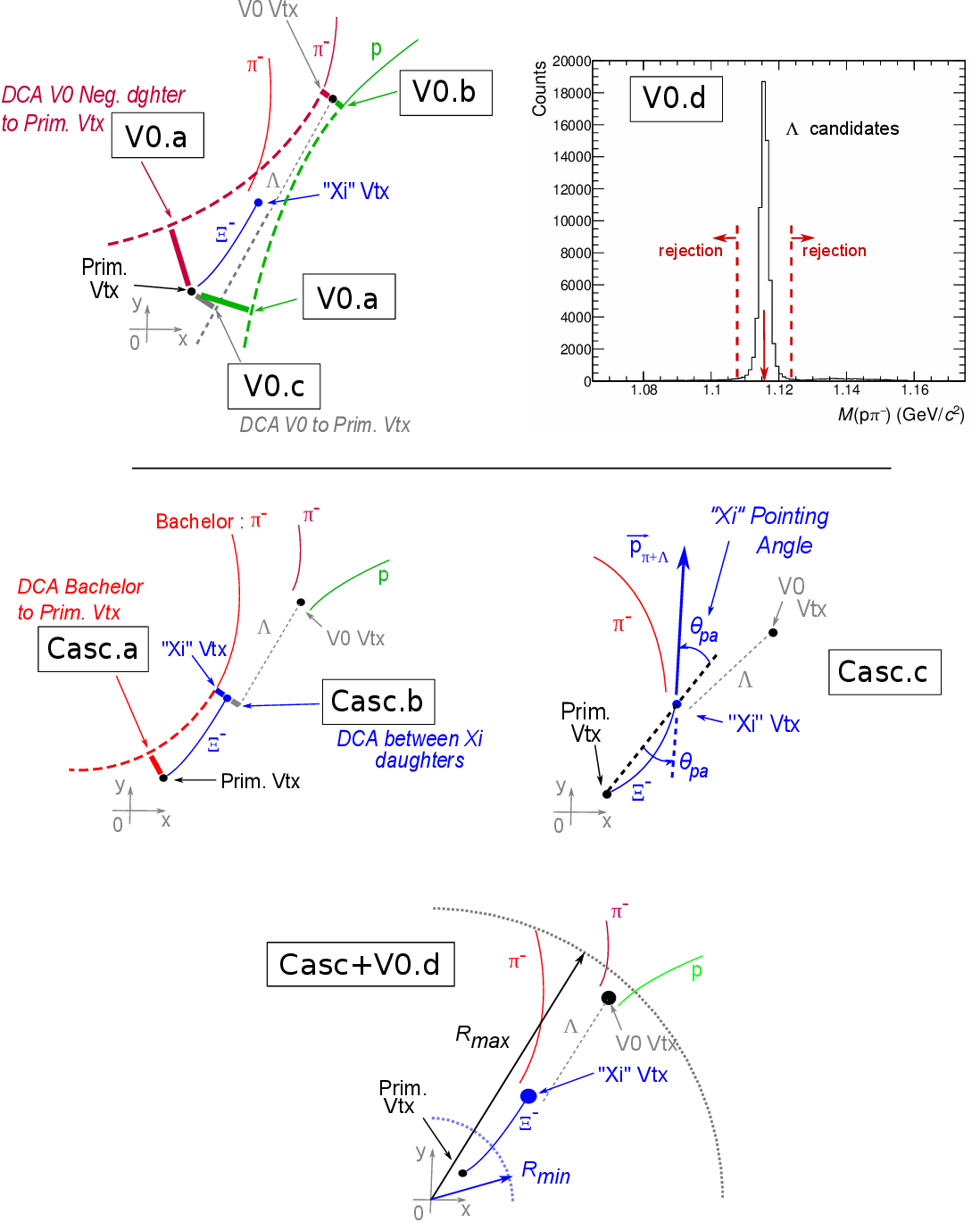}
\caption{Sketches illustrating the various selections used for the topological reconstruction of the multistrange baryons \xim, \xip, \omm, \omp.}\label{fig:SketchTopologicalSelections}
\end{figure}

\begin{table}[!p]
    \centering
    \caption{Summary of the topological and track selections and associated cut values used in the reconstruction of \Xis and \Oms in pp events at \sppt{13}. The \textit{competing mass rejection} refers to the removal of the background contamination from the other cascade species hypothesis, \emph{e.g.} here the removal of actual $\Xis$ contaminants out of the $\Oms$ invariant-mass distribution.}\label{tab:CascadeSelections}
    \begin{tabular}{c|c|c}
    \noalign{\smallskip}\hline \noalign{\smallskip}
    \bf Candidate variable & Selections \Xis & Selections \Oms \\
    \noalign{\smallskip}\hline \noalign{\smallskip}    
    Cascade rapidity interval & \multicolumn{2}{c}{0 < $y$ < 0.5} \\
    Cascade transverse momentum interval (\gev) & 2.4 < $p_{\rm T}$ < 5 & 1.4 < $p_{\rm T}$ < 5 \\
    Competing mass rejection $|M\left(\Xis\right) - M_{\rm PDG} \left(\Xis\right)|$ (\gevcsq) & - & > 0.008 \\

    \noalign{\smallskip}\hline \noalign{\smallskip}
    \bf Track variable & Selections \Xis & Selections \Oms \\
    \noalign{\smallskip}\hline \noalign{\smallskip}
    Pseudorapidity interval & \multicolumn{2}{c}{$0 < \eta  < +0.8$} \\
    Number of crossed TPC readout rows & \multicolumn{2}{c}{ > 70, out of 159} \\
    $\Nsigma^{\rm TPC}$ & \multicolumn{2}{c}{< 3} \\
    \multirow{2}{*}{Out-of-bunch pile-up rejection} & \multicolumn{2}{c}{at least one track with} \\
    & \multicolumn{2}{c}{ITS refit or hit in TOF} \\
    \multirow{2}{*}{Causality check} & \multicolumn{2}{c}{No attached ITS cluster} \\
    & \multicolumn{2}{c}{before the decay point} \\
    \noalign{\smallskip}\hline \noalign{\smallskip}
    \bf Topological variable & Selections \Xis & Selections \Oms \\
    \noalign{\smallskip}\hline \noalign{\smallskip}
    
    \multicolumn{3}{l}{$\mathbf{V^{0}}$} \\
    $\rm V^{0}$ decay radius (cm) & > 1.2 & > 1.1\\
    $\rm V^{0}$ cosine of pointing angle & \multicolumn{2}{c}{> 0.97}\\
    |$M$($\rm V^{0}$) $-$ $M_{\rm PDG} (\plam)$| (\gevcsq) & \multicolumn{2}{c}{< 0.008} \\
    DCA proton to primary vertex (cm) & \multicolumn{2}{c}{> 0.03} \\
    DCA pion to primary vertex (cm) & \multicolumn{2}{c}{> 0.04} \\
    DCA $\rm V^{0}$ to primary vertex (cm) & \multicolumn{2}{c}{> 0.06} \\
    DCA between $\rm V^{0}$ daughters ($\sigma$) & \multicolumn{2}{c}{< 1.5} \\
    \noalign{\smallskip}\hline \noalign{\smallskip}
    
    \multicolumn{3}{l}{\textbf{Cascade}} \\
    Cascade decay radius (cm) & > 0.6 & > 0.5 \\
    Cascade lifetime (cm) & \multicolumn{2}{c}{< 3 $\times$ $c\tau$}\\
    DCA bachelor to primary vertex (cm) & \multicolumn{2}{c}{> 0.04} \\
    DCA between cascade daughters ($\sigma$) & \multicolumn{2}{c}{< 1.3} \\
    Cascade cosine of pointing angle & \multicolumn{2}{c}{> 0.998} \\
    Bachelor--proton pointing angle (rad) & \multicolumn{2}{c}{> 0.04} \\
    
    \noalign{\smallskip}\hline \noalign{\smallskip}
    \end{tabular}
\end{table}

The detection of multistrange baryons relies on the reconstruction of such cascade and $\rm V^{0}$ decay topologies (see Fig.~\ref{fig:SketchTopologicalSelections}).
It is performed at midrapidity ($|y| < 0.5$, further restricted to $0 < y < 0.5$ for reasons discussed below) where rapidity is defined as $y = \frac{1}{2} \ln[(E+p_z)/(E-p_z)]$. Here $E$ is the particle energy and $p_z$ its momentum component along the beam axis. In such a midrapidity window, the production of given (anti)hadrons remains essentially homogeneous, this ensures uniform detector acceptance and stable kinematic conditions. 
The reconstruction uses the standard ALICE weak-decay finder~\cite{alicecollaborationPerformanceALICEExperiment2014}, which starts with the formation of $\rm V^{0}$ candidates. Table~\ref{tab:CascadeSelections} summarises the final selections used in the present analysis. The first step consists of identifying secondary tracks that do not originate from the interaction point. They are tagged as such if the distance of closest approach (DCA) between the considered track and the primary vertex exceeds a critical value (Fig.\ref{fig:SketchTopologicalSelections}, V0.a). The second step aims at forming pairs of secondary tracks of opposite charge, characterised by opposite curvatures; by imposing that the DCA between the two tracks is small, the pairs originating from the same decay point are retained (Fig.\ref{fig:SketchTopologicalSelections}, V0.b). The secondary vertex is then positioned on the segment defined by the previous DCA, weighted by the track quality. The two daughter tracks are then propagated from their initial position (the point of closest approach to the primary vertex) to the secondary decay point. This allows one to calculate all the kinematic quantities of the $\rm V^{0}$, among them its momentum, which is equal to the momentum sum of the positively and negatively charged particles at the secondary vertex, due to momentum conservation. Selection criteria on parameters such as the pointing angle of the $\rm V^{0}$ momentum vector with respect to the primary vertex and the decay radius are applied to suppress combinatorial background.

From the sample of $\rm V^{0}$ candidates, only those compatible with \Xis or \Oms decays are considered. Cascade reconstruction requires a secondary $\rm V^{0}$ consistent with a \plam or \palam. Primary and secondary $\rm V^{0}$s are separated using their pointing direction in the laboratory frame: the straight-line trajectory of the neutral \plam (or \palam) allows calculating the DCA to the interaction point. Primary $\rm V^{0}$s, characterized by vanishing DCA, are rejected (Fig.\ref{fig:SketchTopologicalSelections}, V0.c). The invariant mass is then computed under the \plam or \palam hypothesis, assigning the proton–pion mass combination accordingly (Fig.\ref{fig:SketchTopologicalSelections}, V0.d). True \plam or \palam candidates have reconstructed masses within a few \mevcsq of the nominal value, $M\left(\plam\right)$ = 1.115683 \gevcsq, while misidentified daughters shift the mass outside the selection window, ensuring clean identification.

Cascade candidates are formed by combining a selected \plam (or \palam) with a remaining secondary track playing the role of the bachelor particle (Fig.\ref{fig:SketchTopologicalSelections}, Casc.a). Similarly to $\rm V^{0}$s, the DCA between the $\rm V^{0}$ and the bachelor is required to be small (Fig.\ref{fig:SketchTopologicalSelections}, Casc.b), and the pointing angle between the candidate momentum and the line connecting the primary and secondary vertices must be small for cascade candidates originating from the primary vertex (Fig.\ref{fig:SketchTopologicalSelections}, Casc.c). Additional selections on the decay radius and on the proper lifetime ensure that candidates decaying too close or too far from the interaction point are rejected, guaranteeing a well-defined secondary vertex and reliable tracking of the decay products (Fig.\ref{fig:SketchTopologicalSelections}, Casc+V0.d). To suppress background from mis-reconstructed $\rm V^{0}$s, an additional pointing angle constraint is applied between the bachelor and the proton/antiproton track. This removes cases where a nearby track is mistakenly associated with the proton/antiproton from a \plam/\palam decay to form a $\rm V^{0}$. The remaining \ppim/\ppip daughter is most likely combined with the previous ill-formed $\rm V^{0}$, playing the role of the bachelor particle of a cascade decay. To avoid large contamination from erroneously identified decay daughters, \Oms candidates whose invariant mass computed under the assumption of a \Xis mass hypothesis falls within a window of a few \mevcsq around the nominal \Xis mass value are rejected. To ensure unbiased momentum estimates and thus guarantee stable mass measurements, only cascade candidates reconstructed at intermediate transverse momentum (typically $1.4 < p_{\rm T} < 5~\gev$, with species-dependent boundaries reflecting the different decay kinematics, see Table~\ref{tab:CascadeSelections}) and with small opening angles between the $\rm V^{0}$ and the bachelor particle are retained for the mass determination.

\begin{figure}[t]
\begin{center}
\includegraphics[width=0.8\linewidth,clip,trim={0 0.cm 0 0}]{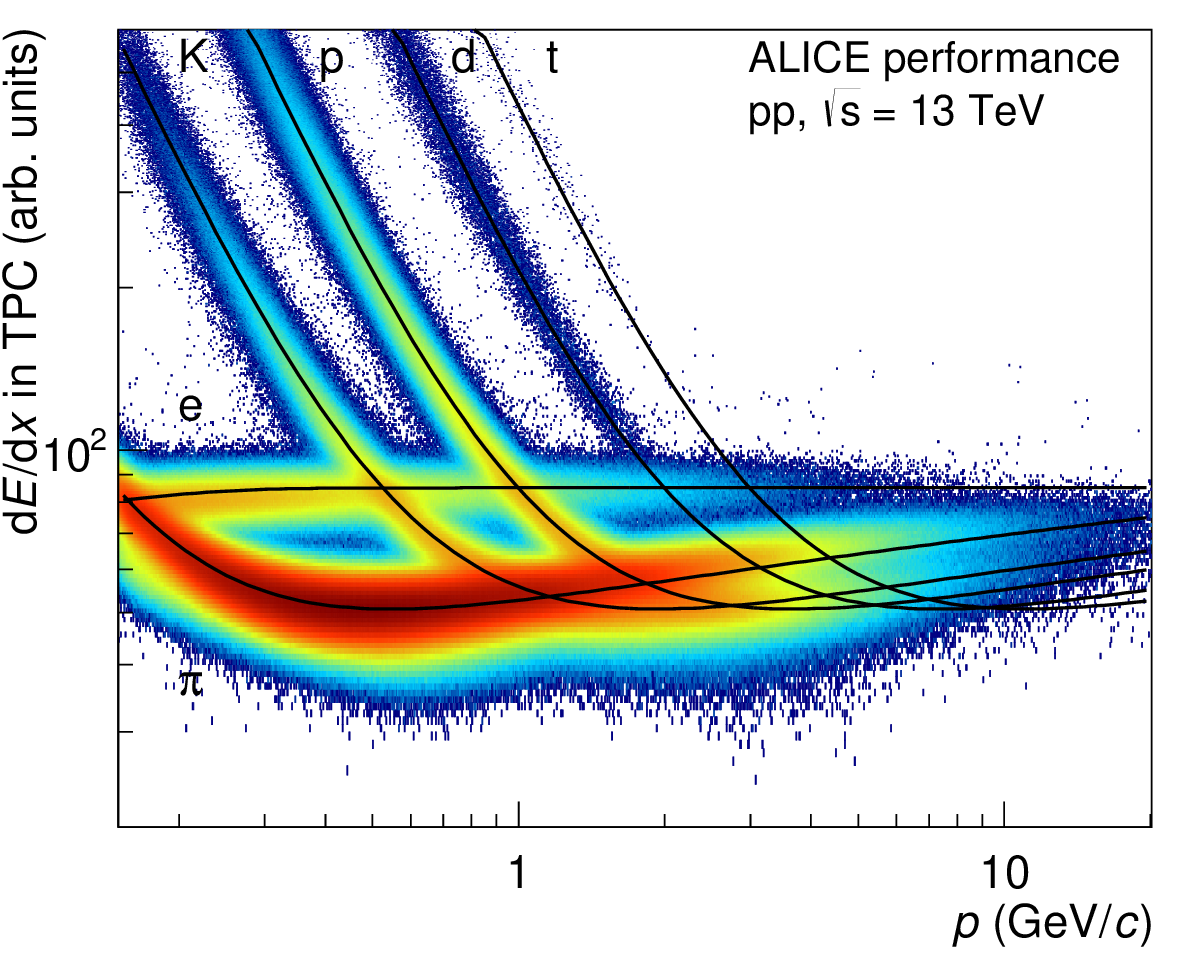}
\end{center}
\caption{TPC \dEdx distributions for charged particles in pp collisions at $\sqrt{s}$ = 13 TeV, overlaid to the expected energy loss per unit of path length given by the Bethe-Bloch formula.} \label{fig:TPCPID} 
\end{figure}

The mother particle and its decay products are required to remain within the central-barrel region, ensuring that the entire decay topology is reconstructed within a fiducial volume where tracking performance is optimal. Furthermore, all decay products are required to lie on the same positive $z$-side of the detector (positive rapidity), where the detector calibration is more stable. Additionally, tracks sharing clusters or exhibiting a kink topology are excluded to avoid bias in the momentum measurement. Particle identification is performed for each daughter track by requiring that the measured specific energy loss ($\mathrm{d}E/\mathrm{d}x$) in the TPC is consistent with the expected value within a defined range of $n_{\sigma}$, where $\sigma$ denotes the TPC energy-loss resolution (see Fig.~\ref{fig:TPCPID}). Tracks must be associated with at least 70 pad rows in the TPC (out of 159 in total) and must have passed successfully the TPC track refit at the end of the reconstruction process, to guarantee both accurate momentum reconstruction and stable particle identification. To further remove the contribution from out-of-bunch pileup events, \textit{i.e.} pile-up from collisions occurring in neighbouring LHC bunch crossings, at least one of the daughter tracks is required to have a hit in the two innermost silicon pixel detectors of the ITS (or be successfully refitted within the ITS) or to be matched to a hit in the TOF detector. The SPD requirement exploits the fast readout time of the silicon pixels, which limits the accepted tracks to those produced within $\pm$300~ns of the triggered bunch crossing (\textit{i.e.} $\pm$12 bunch crossings). The TOF requirement complements this by exploiting the precise timing information of the detector (50 ps) to identify the bunch crossing from which the particle originates. To remove any bias in the invariant mass, a causality requirement is applied: daughter tracks updated with an ITS cluster below the position of the production vertex by more than $1 \sigma$ are discarded; $\sigma$ here refers to the spatial resolution on the vertex. Finally, since the resolution of the decay vertex degrades near an ITS layer, $\rm V^{0}$ and cascade candidates decaying in close proximity to an ITS layer are excluded from the analysis.

\subsection{Energy-loss corrections}\label{ssec:EnergyLossCorrections}

In the standard ALICE framework, tracks are propagated to their DCA to the primary vertex using energy-loss corrections that assume the most probable mass, estimated from a preliminary evaluation of the energy loss in the TPC during event reconstruction~\cite{alicecollaborationPerformanceALICEExperiment2014}. For secondary tracks, this introduces a momentum bias that grows with the distance between the primary and decay vertices, as the energy losses accumulated when propagating from the primary to the decay vertex are not re-applied during the reconstruction of the decay topology. Moreover, the energy-loss corrections assume a uniform material composition regardless of the actual material traversed, and rely on a mass hypothesis that does not necessarily coincide with the true particle mass.
While these approximations are negligible for most analyses, they are critical for the accuracy of the present measurement: the induced mass bias is of $\mathcal{O}(\mevcsq)$, which is one order of magnitude larger than the target precision of this measurement $\mathcal{O}(100~\kevcsq)$.

\begin{figure}[t]
\begin{center}
\includegraphics[width=0.8\linewidth,clip,trim={0 0.cm 0 0}]{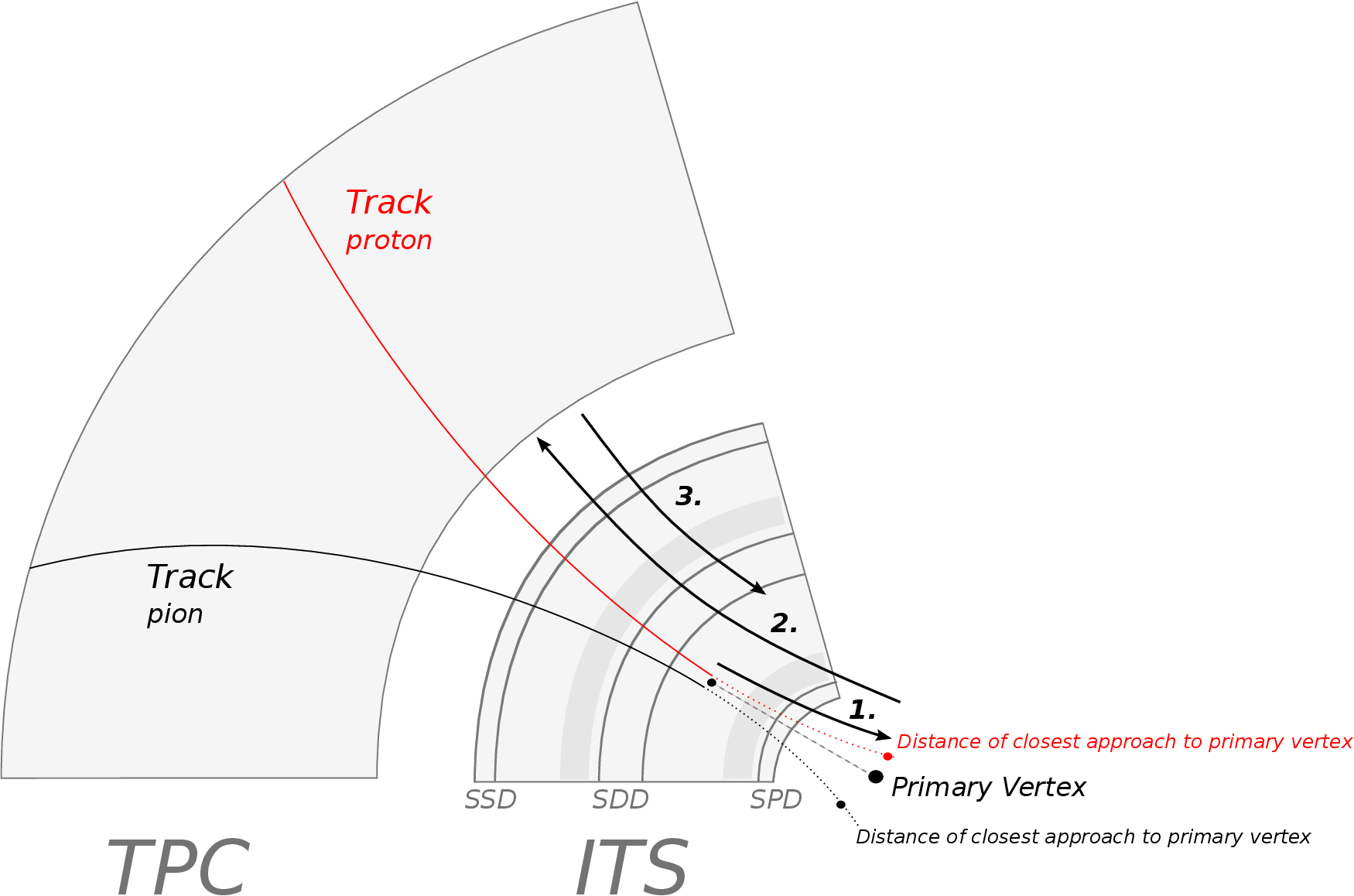}\label{fig:SchemeRetroCorrection}
\end{center}
\caption{Sketch illustrating the energy-loss corrections applied on the proton daughter of a \plam.}
\end{figure}

To address this, a two-stage correction procedure was developed, as illustrated in Fig.~\ref{fig:SchemeRetroCorrection}. The first correction stage consists of two propagation steps (Fig.~\ref{fig:SchemeRetroCorrection}, labels 1 and 2). In the first stage, each daughter track is propagated from the decay vertex to its DCA to the primary vertex (Fig.\ref{fig:SchemeRetroCorrection}, 1.) without applying energy-loss corrections, effectively undoing the inward propagation performed during the standard reconstruction and recovering the track parameters as they were prior to the V$^{0}$ and cascade finding. The track is then propagated to the inner wall of the TPC under the same conditions as the standard reconstruction -- same mass hypothesis and same material assumptions -- such that the energy-loss correction previously applied during the standard reconstruction is exactly removed (Fig.\ref{fig:SchemeRetroCorrection}, 2.). Since the dominant contribution to the material budget comes from the ITS, the residual effect of the incorrect energy-loss corrections within the TPC volume is negligible to first order, as verified by extending the propagation to the TPC outer wall and observing no significant change in the results.
In the second stage, the energy-loss corrections are re-applied from the TPC inner wall to the decay vertex (Fig.\ref{fig:SchemeRetroCorrection}, 3.) using the correct mass hypothesis for each particle species together with the actual material composition of the detector traversed along the trajectory, rather than the uniform approximations of the standard framework. This ensures that the track parameters at the decay vertex accurately reflect the true momentum of each daughter particle.
This procedure was validated using MC simulations, where the input masses are known by construction. 

\subsection{Mass extraction procedure}\label{ssec:MassExtraction}

Figure~\ref{fig:InvMass} presents examples of invariant-mass distributions of the $\Xi^{-}$, $\overline{\Xi}^+$, $\Omega^{-}$, $\overline{\Omega}^+$ obtained in pp collisions at $\sqrt{s}\,=\,13\,$TeV. To isolate the signal from the background, a fit of the invariant-mass distribution is performed using a sum of two functions: one for modelling the signal peak and another for describing the background. Due to the detector smearing, the center of the invariant-mass distribution exhibits a quasi-Gaussian shape but with non-Gaussian tails. Since the resolution of the transverse momentum ($p_\mathrm{T}$, defined as the particle momentum perpendicular to the beam axis) depends on $p_\mathrm{T}$ itself, the invariant-mass distribution is built from reconstructed candidates with different resolutions, which affects the width of the invariant-mass peak. A pseudo-Gaussian, defined as the sum of three Gaussians sharing a common mean but different widths, makes it possible to capture the different widths in a satisfactory manner. 
An asymmetric function, such as the Bukin function~\cite{bukin2007fittingfunctionasymmetricpeaks}, also appears to be a reasonable choice. For the background, an exponential and a linear function were considered. All the combinations between these two pairs of functional forms were tested; the sum of three Gaussians and an exponential function, which yields the best reduced $\chi^{2}$, is adopted as the default fit model, while the other combinations of peak and background functions are used for the study of the systematic uncertainties. In all cases, the fitting procedure is performed with a binned maximum (log-)likelihood method. The particle mass is obtained from the position of the maximum of the fitted signal function, with its statistical uncertainty determined from the fit, while its width provides an estimate of the experimental resolution. The signal and background yields are evaluated from the fit functions.

In total, $15.3 \times 10^{3}\ \Xi^{-}$, $14.8 \times 10^{3}\ \overline{\Xi}^{+}$, $10.1 \times 10^{3}\ \Omega^{-}$ and $9.9 \times 10^{3}\ \overline{\Omega}^{+}$ baryons were selected with a low level of background, with purities reaching 96\% and 91\%, respectively. The signals of $\Xi^{\pm}$ and $\Omega^{\pm}$ are greater by a factor 6 and 100 than the sample exploited in previous mass measurements.
To account for potential biases in the reconstructed mass originating from the data processing chain, event selection, or fitting procedure, a residual mass offset correction is applied. Although the functional forms used for peak and background fitting yield statistically robust mass extractions, a systematic shift is observed between the extracted mass in data and the known PDG values. To correct for this, MC simulations are employed in which the injected mass matches the PDG value by construction. The reconstructed mass in MC is then compared to the input value to determine a correction offset, which is subsequently applied to the measured mass in data. Since this procedure relies on accurate simulation of the detector response, the MC \pt spectra are reweighted to reproduce those observed in data. This ensures that kinematic effects, which could otherwise skew the offset determination, are mitigated. The same correction strategy is applied symmetrically to both particle and antiparticle masses, and is thus consistently propagated to the mass difference. The validity of this correction procedure was verified using \pkazero, $\plam$, and $\palam$ samples, as discussed in the dedicated validation study below.

\begin{figure}[p]
\centering
\includegraphics[angle=90, height=0.92\textheight,clip]{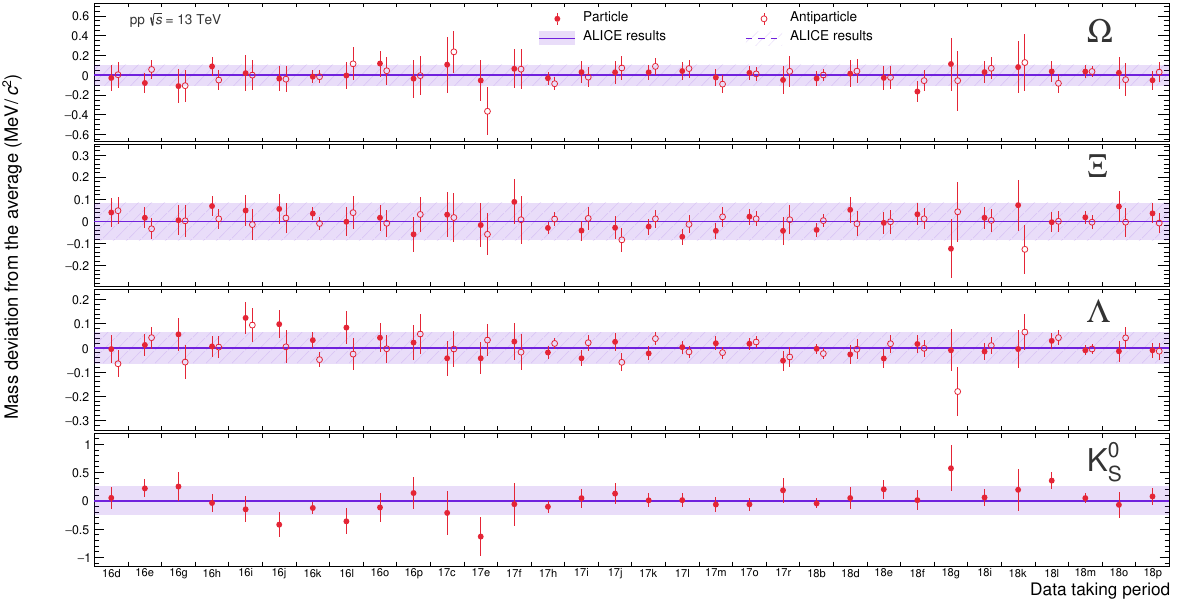}
\caption{Measurements of the mass of \pkazero, $\plam$, $\Xis$ and $\Oms$ with respect to the average value as a function of the data-taking period. Horizontal bands represent the total uncertainties obtained from the combined analysis of all data-taking periods for particles (solid lines) and antiparticles (dashed lines).}\label{fig:MassVsPeriods}
\end{figure}

\subsection{Determination of systematic uncertainties}\label{ssec:SystUncert}

The main sources of systematic uncertainty originate -- directly or indirectly -- from the candidate selections, the finite accuracy on the magnetic-field map, the energy-loss corrections, and the detector calibration, contributing 70 to 90 \kevcsq to the total systematic uncertainty of 78 to 102 \kevcsq. To estimate the effect of the candidate selections, the measurements were performed using 5 000 randomly generated sets of selection criteria. The number of sets was determined to be large enough to ensure stable and statistically robust results. Each selection was varied independently and randomly within a range chosen to induce at least a 10\% variation in the raw signal yield -- a threshold that ensures sensitivity to systematic effects without compromising the statistical significance of the variation. The final results correspond to the average over all the sets; the standard deviation over the 5000 mean mass values obtained is taken as an estimate of the systematic uncertainty. 

To ensure the stability of the results over time, measurements were repeated across 34 distinct data-taking periods from 2016 to 2018, with the L3 solenoid magnet operated in both field configurations ($B=+0.5$ T and $B=-0.5$ T). In this context, a data-taking period corresponds to a time interval during which the accelerator and detector configurations remain unchanged; a new period is started whenever significant changes occur, such as switching the magnetic field polarity, changes in inactive detector regions, variations in instantaneous luminosity, or beam-induced background. As shown in Fig.~\ref{fig:MassVsPeriods}, the results are consistent throughout the entire dataset, demonstrating temporal stability and excellent agreement between measurements taken with opposite magnetic field polarities. 

The magnetic field within the L3 magnet was determined through detailed Hall probe mapping with an accuracy of 0.04\%~\cite{shahoyanSummaryL3Magnet2007}. Its influence on the measurement was assessed by scaling the transverse momenta of the decay daughters accordingly, and the resulting deviation is quoted as a systematic uncertainty. The observed maximum deviation provides the associated systematic uncertainty. The material budget in the central region of the ALICE detector was determined with an accuracy of 2.5\%~\cite{alicecollaborationValidationALICEMaterial2022,alicecollaborationDatadrivenPrecisionDetermination2023}. To assess its impact on the measurement, the full analysis chain is repeated using Monte Carlo simulations in which the material budget is varied within this uncertainty. The resulting deviations from the baseline results obtained with the nominal material budget are assigned as the corresponding systematic uncertainty. 

The measurements were also studied as a function of spatial regions in the ALICE detector, momenta, opening angles, rapidity intervals, and multiplicity classes. Any residual dispersion -- defined as the standard deviation of the results after accounting for all known sources of systematic uncertainty -- is indicative of remaining systematic effects related to the detector calibration. If the observed dispersion exceeds the total expected uncertainty, it is taken as an additional residual systematic uncertainty. This residual is dominated by mass variations with azimuth and is added in quadrature to the other contributions. All other dependencies show negligible residual dispersion compared to the azimuthal variation.

The effect of the transverse-momentum and opening-angle requirements was studied by loosening and tightening the corresponding selection criteria. The impact of the mass-extraction procedure was assessed by varying the functional forms used to describe the signal (symmetric and asymmetric shapes) and background, as well as by changing the fit range and invariant-mass binning. As an independent cross-check, the results were also compared with the mean values extracted directly from the invariant-mass distributions, with no systematic variations observed. The uncertainty associated with the finite precision on the decay-daughter masses was quantified by randomly varying each daughter mass 20 000 times according to a Gaussian distribution centred on the PDG value with its quoted uncertainty as the standard deviation. Possible pile-up effects were probed by removing the pile-up rejection requirement altogether and by extending it from one to two decay daughters. Finally, the MC mass correction can only be as precise as the extracted mass value in MC, which is constrained by the size of the simulated data sample. The uncertainty in the MC mass offset correction was taken to be the statistical precision of the extracted mass value in simulated samples. 
The combined systematic uncertainty from variations of the transverse-momentum and opening-angle requirements, mass-extraction procedure, decay-daughter mass uncertainties, pile-up treatment, and MC mass-offset correction remains small and does not exceed 25~\kevcsq.

In contrast to the absolute mass measurements, most systematic effects cancel in the mass differences due to their symmetric impact on particles and antiparticles. Specifically, uncertainties related to the magnetic field map, material budget, out-of-bunch pile-up rejection, and the precision of reference PDG masses affect particles and antiparticles similarly and lead to a negligible effect in the mass-difference measurement. Residual systematic uncertainties remain from sources that introduce charge-dependent effects. These primarily include topological selection criteria, which can lead to differences in reconstruction efficiency or background characteristics between particles and antiparticles, and momentum calibration, where imperfect charge symmetry in the detector response can introduce biases. Minor contributions also arise from the fitting procedure; although these do not significantly impact the results, they are included for completeness.

Tables~\ref{tab:SystMassXi},~\ref{tab:SystMassOmega} and~\ref{tab:SystMassDiff} summarise the systematic uncertainties in the present analysis.

\begin{table}[p]
    \centering
        \caption{Statistical and systematic uncertainties on the mass of $\Xi^{-}$ and $\overline{\Xi}^{+}$. The total systematic uncertainty is obtained adding each contribution in quadrature, assuming all sources to be uncorrelated.}\label{tab:SystMassXi}
    \begin{tabular}{>{\raggedright\arraybackslash}p{6.2cm} 
                >{\centering\arraybackslash}p{3cm} 
                >{\centering\arraybackslash}p{3cm}}
    \noalign{\smallskip}\hline \noalign{\smallskip}
    \bf & \multicolumn{2}{c}{Uncertainties on the measured mass (\mevcsq)} \\
    \bf Sources & $\Xi^{-}$ & $\overline{\Xi}^{+}$ \\
    \noalign{\smallskip}\hline \noalign{\smallskip}
    \bf Statistical & \bf 0.026 & \bf 0.024 \\
    \noalign{\smallskip}\hline \noalign{\smallskip}
    \bf Systematic (total) & \bf 0.078 & \bf 0.083\\
    \quad Candidate selections (incl. topology) & 0.024 & 0.028\\
    \quad Detector calibration & 0.033 & 0.033 \\
    \quad \pt selections & 0.014 & 0.014 \\
    \quad Opening-angle selections & negligible & negligible  \\
    \quad Magnetic field & 0.023 & 0.028 \\
    \quad Material budget & 0.019 & 0.019 \\
    \quad Mass-extraction procedure & 0.010 & 0.010\\
    \quad Out-of-bunch pile-up rejection & 0.006 & 0.006\\
    \quad Precision on the PDG mass & 0.011 & 0.011 \\
    \quad MC mass offset & 0.055 & 0.058 \\
    \noalign{\smallskip}\hline \noalign{\smallskip}
    \end{tabular}
\end{table}

\begin{table}[p]
    \centering
        \caption{Statistical and systematic uncertainties on the mass of $\Omega^{-}$ and $\overline{\Omega}^{+}$. The total systematic uncertainty is obtained adding each contribution in quadrature, assuming all sources to be uncorrelated.}\label{tab:SystMassOmega}
    \begin{tabular}{>{\raggedright\arraybackslash}p{6.2cm} 
                >{\centering\arraybackslash}p{3cm} 
                >{\centering\arraybackslash}p{3cm}}
    \noalign{\smallskip}\hline \noalign{\smallskip}
    \bf & \multicolumn{2}{c}{Uncertainties on the measured mass (\mevcsq)} \\
    \bf Sources & $\Omega^{-}$ & $\overline{\Omega}^{+}$ \\
    \noalign{\smallskip}\hline \noalign{\smallskip}
    \bf Statistical & \bf 0.033 & \bf 0.034 \\
    \noalign{\smallskip}\hline \noalign{\smallskip}
    \bf Systematic (total) & \bf 0.102 & \bf 0.102\\
    \quad Candidate selections (incl. topology) & 0.027 & 0.035\\
    \quad Detector calibration & 0.089 & 0.086 \\
    \quad \pt selections & 0.005 & 0.005 \\
    \quad Opening-angle selections & negligible & negligible  \\
    \quad Magnetic field & 0.026 & 0.027 \\
    \quad Material budget & 0.012 & 0.012 \\
    \quad Mass-extraction procedure & 0.008 & 0.008\\
    \quad Out-of-bunch pile-up rejection & 0.004 & 0.003\\
    \quad Precision on the PDG mass & 0.018 & 0.018 \\
    \quad MC mass offset & 0.021 & 0.019 \\
    \noalign{\smallskip}\hline \noalign{\smallskip}
    \end{tabular}
\end{table}

\begin{table}[p]
    \centering
    \caption{Statistical and systematic uncertainties on the relative mass difference between $\Xi^{-}$ and $\overline{\Xi}^{+}$, and between $\Omega^{-}$ and $\overline{\Omega}^{+}$. The total systematic uncertainty is obtained adding each contribution in quadrature, assuming all sources to be uncorrelated.}\label{tab:SystMassDiff}
        \begin{tabular}{>{\raggedright\arraybackslash}p{6.2cm} 
                >{\centering\arraybackslash}p{3cm} 
                >{\centering\arraybackslash}p{3cm}}
    \noalign{\smallskip}\hline \noalign{\smallskip}
    \bf & \multicolumn{2}{c}{Uncertainties on the measured} \\
    \bf & \multicolumn{2}{c}{relative mass difference ($\times 10^{-5}$)} \\
    \bf Sources & $\Xi$ & $\Omega$ \\
    \noalign{\smallskip}\hline \noalign{\smallskip}
    \bf Statistical & \bf 2.77 & \bf 2.98 \\
    \noalign{\smallskip}\hline \noalign{\smallskip}
    \bf Systematic (total) & \bf 5.60 & \bf 3.32\\
    \quad Candidate selections (incl. topology) & 3.54 & 2.83\\
    \quad Detector calibration & negligible & negligible \\
    \quad \pt selections & negligible & negligible \\
    \quad Opening-angle selections & negligible & negligible \\
    \quad Magnetic field & negligible & negligible \\
    \quad Material budget & negligible & negligible \\
    \quad Mass-extraction procedure & 0.88 & 0.32\\
    \quad Out-of-bunch pile-up rejection & negligible & negligible\\
    \quad Precision on the PDG mass & negligible & negligible \\
    \quad MC mass offset & 4.27 & 1.69 \\
    \noalign{\smallskip}\hline \noalign{\smallskip}
    \end{tabular}
\end{table}

To ensure that no significant systematic bias remains, the well-known \pkazero\ ($\rm d\bar{s}$) and \plam\ ($\rm uds$) masses, along with the mass difference between \plam\ and \palam~\cite{pdgParticlePhysics2024}, are used as calibration benchmarks. These particles have well-established properties and decay via topologically similar channels: long-lived, neutral particles that decay on average several centimeters away from the primary collision point into two oppositely charged daughters. These control channels are reconstructed using the same tracking and vertexing algorithms as those used for the multistrange baryon candidates, ensuring consistent treatment. 
The resulting masses from the ALICE experiment, presented in Table~\ref{tab:FinalResultsV0s}, are consistent with the PDG masses and determined here with an absolute uncertainty of 250 \kevcsq for \pkazero and 70 \kevcsq for \plam or \palam, each largely dominated by the systematic uncertainties. 
The three masses of these single-strange hadrons are thus determined with a similar absolute mass scale as for cascades, the masses fall within $2\sigma$ of the respective PDG values; the relative mass difference between \plam and \palam is compatible with 0 and reaches a relative accuracy of $2.33 \times 10^{-5}$.

Residual offsets of the \pkazero\ and \plam\ mass values with respect to the PDG can arise from discrepancies in the detector description between data and MC, which would affect the reconstructed momenta of all particles. To quantify the impact of such offsets on the cascade mass measurements, a dedicated propagation study was performed in data by varying the momenta of all decay daughters simultaneously by a common random scaling factor within $\pm$1\% drawn from a uniform distribution, repeated 5000 times. This establishes a linear relationship between the \pkazero\ and \plam\ mass shifts and those of the cascade masses:
\begin{align*} 
\Delta M_{\Xi^{-}} &= a_{\Xi^{-}} \cdot \Delta M_{K^0_S} + b_{\Xi^{-}} \cdot \Delta M_{\Lambda} \\ 
\Delta M_{\bar{\Xi}^{+}} &= a_{\bar{\Xi}^{+}} \cdot \Delta M_{K^0_S} + b_{\bar{\Xi}^{+}} \cdot \Delta M_{\bar{\Lambda}} \\
\Delta M_{\Omega^{-}} &= a_{\Omega^{-}} \cdot \Delta M_{K^0_S} + b_{\Omega^{-}} \cdot \Delta M_{\Lambda} \\
\Delta M_{\bar{\Omega}^{+}} &= a_{\bar{\Omega}^{+}} \cdot \Delta M_{K^0_S} + b_{\bar{\Omega}^{+}} \cdot \Delta M_{\bar{\Lambda}} 
\end{align*}
where $\Delta M$ denotes the mass offset. The values of the coefficients were extracted from a linear fit, giving: $a_{\Xi^{-}} = 0.1738 \pm 0.0001$, $a_{\overline{\Xi}^{+}} = 0.1737 \pm 0.0001$, $a_{\Omega^{-}} = 0.2597 \pm 0.0001$, $a_{\overline{\Omega}^{+}} = 0.2614 \pm 0.0001$, and $b_{\Xi^{-}} = 0.5377 \pm 0.0001$, $b_{\overline{\Xi}^{+}} = 0.5376 \pm 0.0001$, $b_{\Omega^{-}} = 0.6723 \pm 0.0001$, $b_{\overline{\Omega}^{+}} = 0.6719 \pm 0.0001$.

Inserting the observed \pkazero\ and \plam\ mass shifts, the propagated offsets are estimated to be $0.049$~\mevcsq for the \Xis\ and $0.060-0.061$~\mevcsq for the \Oms. These values are well within the systematic uncertainties of the respective measurements, confirming that any residual bias in the mass scale remains within acceptable limits.

\begin{table}[p]
    \caption{Left: final measured masses of \pkazero, \plam, \palam and the relative mass differences between \plam and \palam, with their associated statistical and systematic uncertainties. Right: tabulated values in the~\cite{pdgParticlePhysics2024}, with their associated uncertainties.}\label{tab:FinalResultsV0s}
    \begin{tabular}{cccc|cc}
    \noalign{\smallskip}\hline \noalign{\smallskip}
    \bf Particle & \bf Measured & \multicolumn{2}{c|}{\bf Uncertainty} & \bf PDG & \bf PDG\\
    & \bf mass & \bf stat. & \bf syst. & \bf mass & \bf uncertainty\\
    & (\mevcsq) & (\mevcsq) & (\mevcsq) & (\mevcsq) & (\mevcsq) \\
    \noalign{\smallskip}\hline \noalign{\smallskip}
    \pkazero & 497.604 & 0.035 & 0.254 & 497.611 & 0.013 \\
    \noalign{\smallskip}\hline \noalign{\smallskip}
    \plam & 1115.776 & 0.006 & 0.065 & \multirow{2}*{1115.683} & \multirow{2}*{0.006} \\ 
    \palam & 1115.775 & 0.006 & 0.064 & & \\ 
	\noalign{\smallskip}\hline \noalign{\smallskip}
	\bf Particle & \bf Measured relative & \multicolumn{2}{c|}{\bf Uncertainty} & \bf PDG relative  & \bf PDG\\
    & \bf mass difference & \bf stat. & \bf syst. & \bf mass difference & \bf uncertainty \\
    & ($\times 10^{-5}$) & ($\times 10^{-5}$) & ($\times 10^{-5}$) & ($\times 10^{-5}$) & ($\times 10^{-5}$)\\
    \noalign{\smallskip}\hline \noalign{\smallskip}
    \palam & 0.02 & 0.68 & 2.22 & 0.1 & 1.1 \\
	\noalign{\smallskip}\hline \noalign{\smallskip}
    \end{tabular}
\end{table}

\newpage
\section{The ALICE Collaboration}
\label{app:collab}
\begin{flushleft} 
\small

D.A.H.~Abdallah\,\orcidlink{0000-0003-4768-2718}\,$^{\rm 134}$, 
I.J.~Abualrob\,\orcidlink{0009-0005-3519-5631}\,$^{\rm 112}$, 
S.~Acharya\,\orcidlink{0000-0002-9213-5329}\,$^{\rm 49}$, 
K.~Agarwal\,\orcidlink{0000-0001-5781-3393}\,$^{\rm II,}$$^{\rm 23}$, 
G.~Aglieri Rinella\,\orcidlink{0000-0002-9611-3696}\,$^{\rm 32}$, 
L.~Aglietta\,\orcidlink{0009-0003-0763-6802}\,$^{\rm 24}$, 
N.~Agrawal\,\orcidlink{0000-0003-0348-9836}\,$^{\rm 25}$, 
Z.~Ahammed\,\orcidlink{0000-0001-5241-7412}\,$^{\rm 132}$, 
S.~Ahmad\,\orcidlink{0000-0003-0497-5705}\,$^{\rm 15}$, 
Z.~Akbar$^{\rm 79}$
% ...

\section*{Affiliation Notes}

$^{\rm I}$ Deceased\\
$^{\rm II}$ Also at: INFN Trieste, Trieste, Italy\\
$^{\rm III}$ Also at: Fondazione Bruno Kessler (FBK), Trento, Italy\\
$^{\rm IV}$ Also at: Czech Technical University in Prague, Prague, Czech Republic\\
$^{\rm V}$ Also at: Instituto de Fisica da Universidade de Sao Paulo\\
$^{\rm VI}$ Also at: Dipartimento DET del Politecnico di Torino, Turin, Italy\\
$^{\rm VII}$ Also at: University College of Dublin, Dublin, Ireland\\
$^{\rm VIII}$ Also at: Department of Applied Physics, Aligarh Muslim University, Aligarh, India\\
$^{\rm IX}$ Also at: Institute of Theoretical Physics, University of Wroclaw, Wroclaw, Poland\\
$^{\rm X}$ Also at: Facultad de Ciencias, Universidad Nacional Aut\'{o}noma de M\'{e}xico, Mexico City, Mexico\\

\section*{Collaboration Institutes}

$^{1}$ A.I. Alikhanyan National Science Laboratory (Yerevan Physics Institute) Foundation, Yerevan, Armenia\\
$^{2}$ AGH University of Krakow, Cracow, Poland\\
$^{3}$ Bogolyubov Institute for Theoretical Physics, National Academy of Sciences of Ukraine, Kyiv, Ukraine\\
$^{4}$ Bose Institute, Department of Physics  and Centre for Astroparticle Physics and Space Science (CAPSS), Kolkata, India\\
$^{5}$ California Polytechnic State University, San Luis Obispo, California, United States\\
$^{6}$ Central China Normal University, Wuhan, China\\
$^{7}$ Centro de Aplicaciones Tecnol\'{o}gicas y Desarrollo Nuclear (CEADEN), Havana, Cuba\\
$^{8}$ Centro de Investigaci\'{o}n y de Estudios Avanzados (CINVESTAV), Mexico City and M\'{e}rida, Mexico\\
$^{9}$ Chicago State University, Chicago, Illinois, United States\\
$^{10}$ China Nuclear Data Center, China Institute of Atomic Energy, Beijing, China\\
$^{11}$ China University of Geosciences, Wuhan, China\\
$^{12}$ Chungbuk National University, Cheongju, Republic of Korea\\
$^{13}$ Comenius University Bratislava, Faculty of Mathematics, Physics and Informatics, Bratislava, Slovak Republic\\
$^{14}$ Creighton University, Omaha, Nebraska, United States\\
$^{15}$ Department of Physics, Aligarh Muslim University, Aligarh, India\\
$^{16}$ Department of Physics, Pusan National University, Pusan, Republic of Korea\\
$^{17}$ Department of Physics, Sejong University, Seoul, Republic of Korea\\
$^{18}$ Department of Physics, University of California, Berkeley, California, United States\\
$^{19}$ Department of Physics, University of Oslo, Oslo, Norway\\
$^{20}$ Department of Physics and Technology, University of Bergen, Bergen, Norway\\
$^{21}$ Dipartimento di Fisica, Universit\`{a} di Pavia, Pavia, Italy\\
$^{22}$ Dipartimento di Fisica dell'Universit\`{a} and Sezione INFN, Cagliari, Italy\\
$^{23}$ Dipartimento di Fisica dell'Universit\`{a} and Sezione INFN, Trieste, Italy\\
$^{24}$ Dipartimento di Fisica dell'Universit\`{a} and Sezione INFN, Turin, Italy\\
$^{25}$ Dipartimento di Fisica e Astronomia dell'Universit\`{a} and Sezione INFN, Bologna, Italy\\
$^{26}$ Dipartimento di Fisica e Astronomia dell'Universit\`{a} and Sezione INFN, Catania, Italy\\
$^{27}$ Dipartimento di Fisica e Astronomia dell'Universit\`{a} and Sezione INFN, Padova, Italy\\
$^{28}$ Dipartimento di Fisica `E.R.~Caianiello' dell'Universit\`{a} and Gruppo Collegato INFN, Salerno, Italy\\
$^{29}$ Dipartimento DISAT del Politecnico and Sezione INFN, Turin, Italy\\
$^{30}$ Dipartimento di Scienze MIFT, Universit\`{a} di Messina, Messina, Italy\\
$^{31}$ Dipartimento Interateneo di Fisica `M.~Merlin' and Sezione INFN, Bari, Italy\\
$^{32}$ European Organization for Nuclear Research (CERN), Geneva, Switzerland\\
$^{33}$ Faculty of Electrical Engineering, Mechanical Engineering and Naval Architecture, University of Split, Split, Croatia\\
$^{34}$ Faculty of Nuclear Sciences and Physical Engineering, Czech Technical University in Prague, Prague, Czech Republic\\
$^{35}$ Faculty of Physics, Sofia University, Sofia, Bulgaria\\
$^{36}$ Faculty of Science, P.J.~\v{S}af\'{a}rik University, Ko\v{s}ice, Slovak Republic\\
$^{37}$ Faculty of Technology, Environmental and Social Sciences, Bergen, Norway\\
$^{38}$ Frankfurt Institute for Advanced Studies, Johann Wolfgang Goethe-Universit\"{a}t Frankfurt, Frankfurt, Germany\\
$^{39}$ Fudan University, Shanghai, China\\
$^{40}$ Gauhati University, Department of Physics, Guwahati, India\\
$^{41}$ Helmholtz-Institut f\"{u}r Strahlen- und Kernphysik, Rheinische Friedrich-Wilhelms-Universit\"{a}t Bonn, Bonn, Germany\\
$^{42}$ Helsinki Institute of Physics (HIP), Helsinki, Finland\\
$^{43}$ High Energy Physics Group,  Universidad Aut\'{o}noma de Puebla, Puebla, Mexico\\
$^{44}$ Horia Hulubei National Institute of Physics and Nuclear Engineering, Bucharest, Romania\\
$^{45}$ HUN-REN Wigner Research Centre for Physics, Budapest, Hungary\\
$^{46}$ Indian Institute of Technology Bombay (IIT), Mumbai, India\\
$^{47}$ Indian Institute of Technology Indore, Indore, India\\
$^{48}$ INFN, Laboratori Nazionali di Frascati, Frascati, Italy\\
$^{49}$ INFN, Sezione di Bari, Bari, Italy\\
$^{50}$ INFN, Sezione di Bologna, Bologna, Italy\\
$^{51}$ INFN, Sezione di Cagliari, Cagliari, Italy\\
$^{52}$ INFN, Sezione di Catania, Catania, Italy\\
$^{53}$ INFN, Sezione di Padova, Padova, Italy\\
$^{54}$ INFN, Sezione di Pavia, Pavia, Italy\\
$^{55}$ INFN, Sezione di Torino, Turin, Italy\\
$^{56}$ INFN, Sezione di Trieste, Trieste, Italy\\
$^{57}$ Inha University, Incheon, Republic of Korea\\
$^{58}$ Institute for Gravitational and Subatomic Physics (GRASP), Utrecht University/Nikhef, Utrecht, Netherlands\\
$^{59}$ Institute of Experimental Physics, Slovak Academy of Sciences, Ko\v{s}ice, Slovak Republic\\
$^{60}$ Institute of Physics, Homi Bhabha National Institute, Bhubaneswar, India\\
$^{61}$ Institute of Physics of the Czech Academy of Sciences, Prague, Czech Republic\\
$^{62}$ Institute of Space Science (ISS), Bucharest, Romania\\
$^{63}$ Institut f\"{u}r Kernphysik, Johann Wolfgang Goethe-Universit\"{a}t Frankfurt, Frankfurt, Germany\\
$^{64}$ Instituto de Ciencias Nucleares, Universidad Nacional Aut\'{o}noma de M\'{e}xico, Mexico City, Mexico\\
$^{65}$ Instituto de F\'{i}sica, Universidade Federal do Rio Grande do Sul (UFRGS), Porto Alegre, Brazil\\
$^{66}$ Instituto de F\'{\i}sica, Universidad Nacional Aut\'{o}noma de M\'{e}xico, Mexico City, Mexico\\
$^{67}$ iThemba LABS, National Research Foundation, Somerset West, South Africa\\
$^{68}$ Jeonbuk National University, Jeonju, Republic of Korea\\
$^{69}$ Korea Institute of Science and Technology Information, Daejeon, Republic of Korea\\
$^{70}$ Laboratoire de Physique Subatomique et de Cosmologie, Universit\'{e} Grenoble-Alpes, CNRS-IN2P3, Grenoble, France\\
$^{71}$ Lawrence Berkeley National Laboratory, Berkeley, California, United States\\
$^{72}$ Lund University Department of Physics, Division of Particle Physics, Lund, Sweden\\
$^{73}$ Marietta Blau Institute, Vienna, Austria\\
$^{74}$ Nagasaki Institute of Applied Science, Nagasaki, Japan\\
$^{75}$ Nara Women{'}s University (NWU), Nara, Japan\\
$^{76}$ National Centre for Nuclear Research, Warsaw, Poland\\
$^{77}$ National Institute of Science Education and Research, Homi Bhabha National Institute, Jatni, India\\
$^{78}$ National Nuclear Research Center, Baku, Azerbaijan\\
$^{79}$ National Research and Innovation Agency - BRIN, Jakarta, Indonesia\\
$^{80}$ Niels Bohr Institute, University of Copenhagen, Copenhagen, Denmark\\
$^{81}$ Nikhef, National institute for subatomic physics, Amsterdam, Netherlands\\
$^{82}$ Nuclear Physics Group, STFC Daresbury Laboratory, Daresbury, United Kingdom\\
$^{83}$ Nuclear Physics Institute of the Czech Academy of Sciences, Husinec-\v{R}e\v{z}, Czech Republic\\
$^{84}$ Oak Ridge National Laboratory, Oak Ridge, Tennessee, United States\\
$^{85}$ Ohio State University, Columbus, Ohio, United States\\
$^{86}$ Physics department, Faculty of science, University of Zagreb, Zagreb, Croatia\\
$^{87}$ Physics Department, Panjab University, Chandigarh, India\\
$^{88}$ Physics Department, University of Jammu, Jammu, India\\
$^{89}$ Physics Program and International Institute for Sustainability with Knotted Chiral Meta Matter (WPI-SKCM$^{2}$), Hiroshima University, Hiroshima, Japan\\
$^{90}$ Physikalisches Institut, Eberhard-Karls-Universit\"{a}t T\"{u}bingen, T\"{u}bingen, Germany\\
$^{91}$ Physikalisches Institut, Ruprecht-Karls-Universit\"{a}t Heidelberg, Heidelberg, Germany\\
$^{92}$ Physik Department, Technische Universit\"{a}t M\"{u}nchen, Munich, Germany\\
$^{93}$ Politecnico di Bari and Sezione INFN, Bari, Italy\\
$^{94}$ Research Division and ExtreMe Matter Institute EMMI, GSI Helmholtzzentrum f\"ur Schwerionenforschung GmbH, Darmstadt, Germany\\
$^{95}$ Saga University, Saga, Japan\\
$^{96}$ Saha Institute of Nuclear Physics, Homi Bhabha National Institute, Kolkata, India\\
$^{97}$ School of Physics and Astronomy, University of Birmingham, Birmingham, United Kingdom\\
$^{98}$ Secci\'{o}n F\'{\i}sica, Departamento de Ciencias, Pontificia Universidad Cat\'{o}lica del Per\'{u}, Lima, Peru\\
$^{99}$ SUBATECH, IMT Atlantique, Nantes Universit\'{e}, CNRS-IN2P3, Nantes, France\\
$^{100}$ Sungkyunkwan University, Suwon City, Republic of Korea\\
$^{101}$ Suranaree University of Technology, Nakhon Ratchasima, Thailand\\
$^{102}$ Technical University of Ko\v{s}ice, Ko\v{s}ice, Slovak Republic\\
$^{103}$ The Henryk Niewodniczanski Institute of Nuclear Physics, Polish Academy of Sciences, Cracow, Poland\\
$^{104}$ The University of Texas at Austin, Austin, Texas, United States\\
$^{105}$ Universidad Aut\'{o}noma de Sinaloa, Culiac\'{a}n, Mexico\\
$^{106}$ Universidade de S\~{a}o Paulo (USP), S\~{a}o Paulo, Brazil\\
$^{107}$ Universidade Estadual de Campinas (UNICAMP), Campinas, Brazil\\
$^{108}$ Universidade Federal do ABC, Santo Andre, Brazil\\
$^{109}$ Universitatea Nationala de Stiinta si Tehnologie Politehnica Bucuresti, Bucharest, Romania\\
$^{110}$ University of Cape Town, Cape Town, South Africa\\
$^{111}$ University of Derby, Derby, United Kingdom\\
$^{112}$ University of Houston, Houston, Texas, United States\\
$^{113}$ University of Jyv\"{a}skyl\"{a}, Jyv\"{a}skyl\"{a}, Finland\\
$^{114}$ University of Kansas, Lawrence, Kansas, United States\\
$^{115}$ University of Liverpool, Liverpool, United Kingdom\\
$^{116}$ University of Science and Technology of China, Hefei, China\\
$^{117}$ University of Silesia in Katowice, Katowice, Poland\\
$^{118}$ University of South-Eastern Norway, Kongsberg, Norway\\
$^{119}$ University of Tennessee, Knoxville, Tennessee, United States\\
$^{120}$ University of the Witwatersrand, Johannesburg, South Africa\\
$^{121}$ University of Tokyo, Tokyo, Japan\\
$^{122}$ University of Tsukuba, Tsukuba, Japan\\
$^{123}$ Universit\"{a}t M\"{u}nster, Institut f\"{u}r Kernphysik, M\"{u}nster, Germany\\
$^{124}$ Universit\'{e} Clermont Auvergne, CNRS/IN2P3, LPC, Clermont-Ferrand, France\\
$^{125}$ Universit\'{e} de Lyon, CNRS/IN2P3, Institut de Physique des 2 Infinis de Lyon, Lyon, France\\
$^{126}$ Universit\'{e} de Strasbourg, CNRS, IPHC UMR 7178, F-67000 Strasbourg, France, Strasbourg, France\\
$^{127}$ Universit\'{e} Paris-Saclay, Centre d'Etudes de Saclay (CEA), IRFU, D\'{e}partment de Physique Nucl\'{e}aire (DPhN), Saclay, France\\
$^{128}$ Universit\'{e}  Paris-Saclay, CNRS/IN2P3, IJCLab, Orsay, France\\
$^{129}$ Universit\`{a} degli Studi di Foggia, Foggia, Italy\\
$^{130}$ Universit\`{a} del Piemonte Orientale, Vercelli, Italy\\
$^{131}$ Universit\`{a} di Brescia, Brescia, Italy\\
$^{132}$ Variable Energy Cyclotron Centre, Homi Bhabha National Institute, Kolkata, India\\
$^{133}$ Warsaw University of Technology, Warsaw, Poland\\
$^{134}$ Wayne State University, Detroit, Michigan, United States\\
$^{135}$ Yale University, New Haven, Connecticut, United States\\
$^{136}$ Yildiz Technical University, Istanbul, Turkey\\
$^{137}$ Yonsei University, Seoul, Republic of Korea\\
$^{138}$ Affiliated with an institute formerly covered by a cooperation agreement with CERN\\
$^{139}$ Affiliated with an international laboratory covered by a cooperation agreement with CERN.\\

\end{flushleft} 
  
\end{document}